\documentclass[fp,twocolumn]{jpsj3}
\usepackage{txfonts}
\usepackage{booktabs}
\usepackage[dvipdfmx]{graphicx}
\usepackage{color}
\usepackage{dcolumn}
\usepackage{bm}
\usepackage[version=4]{mhchem}
\usepackage{xspace}
\usepackage{ulem} 
\newcommand{\gpa}{\mathrm{\; GPa}}
\newcommand{\kelvin}{\mathrm{\; K}}
\newcommand{\ev}{\mathrm{\; eV}}
\newcommand{\mev}{\mathrm{\;meV}}
\newcommand{\ute}{\ce{UTe2}\xspace}
\newcommand{\tc}{T_\mathrm{c}}

\title{Electronic Structure of \ute under Pressure}
\author{Makoto Shimizu\thanks{shimizu.makoto.5c@kyoto-u.ac.jp} and Youichi Yanase}
\inst{Department of Physics, Graduate School of Science, Kyoto University, Kyoto 606-8502, Japan}

\abst{
A heavy-fermion paramagnet \ute has been a strong candidate for a spin-triplet superconductor.
Experiments on \ute under pressure have been vigorously conducted, and rich phase diagrams have been suggested.
Multiple superconducting phases exist in the pressure region of $0 \leq P < 1.8 \gpa$,
and an antiferromagnetic ordered state is observed in the high pressure region $P > 1.8 \gpa$.
However, under pressure, the underlying electronic structure in the normal state has not been clarified, although knowledge of electronic structures is essential for studying magnetic and superconducting states.
As an indispensable step toward understanding the phase diagram of \ute, we study the electronic structure under hydrostatic and uniaxial stresses based on the density functional theory with and without employing structural optimization. It is shown that the low-energy band structure and Fermi surfaces are not sensitive to pressure for parameters where itinerant $f$-electrons are not essential. However, we find a significant pressure dependence for a certain Coulomb interaction $U$ of the GGA+$U$ calculation, where the large weight of $f$-electrons appears at the Fermi level. 
An increase in the density of states at the Fermi level is observed under pressure, which is attributed to compressive stress along the [010] crystallographic axis.
}

\begin{document}
\maketitle

\section{Introduction}

\ute is a strong candidate for a spin-triplet superconductor~\cite{Ran2019,Aoki_2022review,Lewin2023},
which has been searched for several decades, and consequently a candidate for a topological superconductor~\cite{Aoki_2022review,Sato2016}. 
Since the discovery of superconductivity~\cite{Ran2019}, 
\ute is expected to be related to ferromagnetic superconductors, UGe$_2$, URhGe and UCoGe, and thus spin-triplet superconductivity mediated by ferromagnetic fluctuations has been discussed.
However, while ferromagnetic fluctuations have been suggested~\cite{Ran2019,Tokunaga2019,Sundar2019}, antiferromagnetic fluctuations are detected by neutron scattering experiments~\cite{Duan2020, Duan2021,Raymond2021,Knafo2021}.
Therefore, the mechanism of superconductivity in UTe$_2$ is elusive, and it is an urgent issue to elucidate superconductivity from a microscopic point of view.

The application of pressure is a promising way not only to search for new phenomena but also to reveal essential aspects of materials
in the strongly correlated region.
Experiments under hydrostatic pressure have been performed on \ute,
and intriguing electronic phase diagrams have been revealed~\cite{Braithwaite2019,Lin2020,Thomas2020,Aoki2020,Knebel2020,Ran2020,Valiska2021,Aoki2021,Kinjo2023,knafo2025incommensurateantiferromagnetismute2pressure}.
Three different superconducting phases are observed at low pressure ($0 \leq P < 1.8\gpa$),
and an antiferromagnetic phase is observed at high pressure ($P > 1.8\gpa$).
The phase diagram implies that the multiple superconducting phases may be affected
by antiferromagnetic fluctuations originating from the antiferromagnetic ordering at high pressure
and possibly by ferromagnetic fluctuations that could originate from the putative ferromagnetic phase at negative pressure.
The rich phase diagram of \ute has attracted much attention, and theoretical studies on multi-component superconducting order parameters under pressure have been conducted~\cite{Ishizuka2021,Kanasugi2022,Chazono2023,Kitamura2023,Hakuno2024,Tei2024}. 
However, essential properties of superconductivity such as the microscopic origin of Cooper pairing, symmetry of order parameter, and possible topological superconductivity are still elusive.
A critical deficiency is the lack of information on the electronic structure under pressure.
The elucidation of the electronic structure will be the basis for the microscopic understanding of magnetism and superconductivity in \ute.

In addition, the electronic structure, magnetic properties, and superconductivity of \ute show strong anisotropy, and therefore uniaxial stress is expected to cause directional effects which would provide rich information. 
Thermodynamic and electrical transport measurements revealed a
decrease in the transition temperature $\tc$ under uniaxial stress along the [100] and [110] axes
but an increase in $\tc$ along the [001] axis~\cite{Girod2022}.
Ultrasound measurements show
that $\tc$ decreases under uniaxial stress along the [100] and [010] axes and increases along the [001] axis~\cite{Theuss2024},  cosistent with Ref.~\cite{Girod2022}.
It is revealed that the transition temperature is more sensitive to stress along [100] and [001] than along [010]. 
Although further experimental studies are desirable, theoretical studies of the effects of uniaxial stress can pave the way for strain engineering of superconductivity and possibly realize a new phase in \ute.

In light of the recent studies, an {\it ab initio} study of the electronic structure in \ute under hydrostatic and uniaxial stresses is helpful and prescient. In this study, we apply the GGA+$U$ method to \ute under pressure. In general, the applicability of the GGA+$U$ method should be verified for each compound. In \ute, the presence of itinerant $f$-electron bands is indicated by specific heat measurements~\cite{Aoki_2022review,Ran2019,Aoki2019}, and the electronic structures predicted by the GGA+$U$ calculation show good agreement with dHvA experiments~\cite{Aoki_dHvA2022,Aoki_dHvA2023,Broyles2023,Eaton2024,Weinberger2024}. Therefore, although correlation-induced quasiparticle renormalization should be evaluated by other methods, the GGA+$U$ method is a reasonable starting point to investigate the low-energy electronic structure of \ute.

In this paper, we start with GGA+$U$ calculations in a high-quality crystal~\cite{Sakai2022} at ambient pressure. 
We obtain a very similar $U$ dependence of electronic structures to that reported for parameters of conventional-quality crystals~\cite{Ishizuka2019}. 
We prepare lattice constants for various pressures by interpolating experimental data and determine atomic coordinates by executing ionic relaxations.
Using the calculated crystal structure, we calculate electronic band structures with variations of $U$ and $P$. 
We also discuss the possibility of a pressure-induced Lifshitz transition, which changes the character of topological superconductivity.
Next, we study the electronic structure under uniaxial stress.
We prepare lattice constants using elastic constants estimated in a resonant ultrasound measurement~\cite{Theuss2024_elasticmoduli} 
and calculate atomic coordinates by performing ionic relaxation.
We show the contrasting pressure dependence of electronic structures depending on the direction of uniaxial stress.

\section{Electronic Structure at Ambient Pressure}

\begin{table}[h!t]
    \centering
    \begin{tabular}{c|c|c|c}
      \toprule
      & $a$ & $b$ & $c$ \\
      \hline
      Lattice constants ($\AA$) & 4.1618 & 6.1355 & 13.9698 \\
      \toprule
      & $z$ of U $4i$ & $y$ of Te $4h$ & $z$ of Te $4j$ \\
      \hline
      Fractional coordinates & 0.13520 & 0.24910 & 0.29780 \\
      \toprule
    \end{tabular}
    \caption{
      Lattice constants and fractional coordinates of a high-quality \ute crystal
      exhibiting superconductivity below $\tc = 2.1\kelvin$ from Refs.~\cite{Haga2022,Sakai2022}.
    }
    \label{tab:ambientpressure_crystalstructure}
\end{table}

In this study, we investigate the electronic structure in high-quality single crystals of \ute~\cite{Haga2022,Sakai2022}
which show the high superconducting critical temperature $\tc = 2.1\kelvin$ at ambient pressure.
Because previous DFT studies~\cite{Ishizuka2019,Harima2020} assumed slightly different crystal structure parameters for low-$T_{\rm c}$ samples prepared by the CVD method, it is reasonable to start from recalculating the electronic structure at ambient pressure.
We use the crystallographic parameters (TABLE~\ref{tab:ambientpressure_crystalstructure})
of the high-quality single crystal~\cite{Haga2022,Sakai2022}.
The crystal structure of \ute belongs to the $Immm$ (No.~71) space group.
The U atoms are located on the $4i$ Wyckoff positions,
and the Te atoms are located on the $4h$ and $4j$ Wyckoff positions.

We perform fully relativistic all-electron density functional theory (DFT) calculations
using the full potential localized orbital (FPLO) basis~\cite{Koepernik1999}
and the generalized gradient approximation (GGA) exchange correlation functional~\cite{Perdew1996}.
We use the around mean-field functional~\cite{Ylvisaker2009} for double-counting corrections.  
We execute GGA+$U$ calculations in a range of $0 \leq U \leq 2 \ev$ on uranium $5f$ orbitals.
We set $J=0$ for simplification,
and thus our $U$ is an effective Coulomb interaction,
that is the Hubbard interaction minus the Hund's coupling.
We adopt a $12 \times 12 \times 12$ $k$-mesh.

\begin{figure}[htb]
  \centering
  \includegraphics[width=\linewidth]{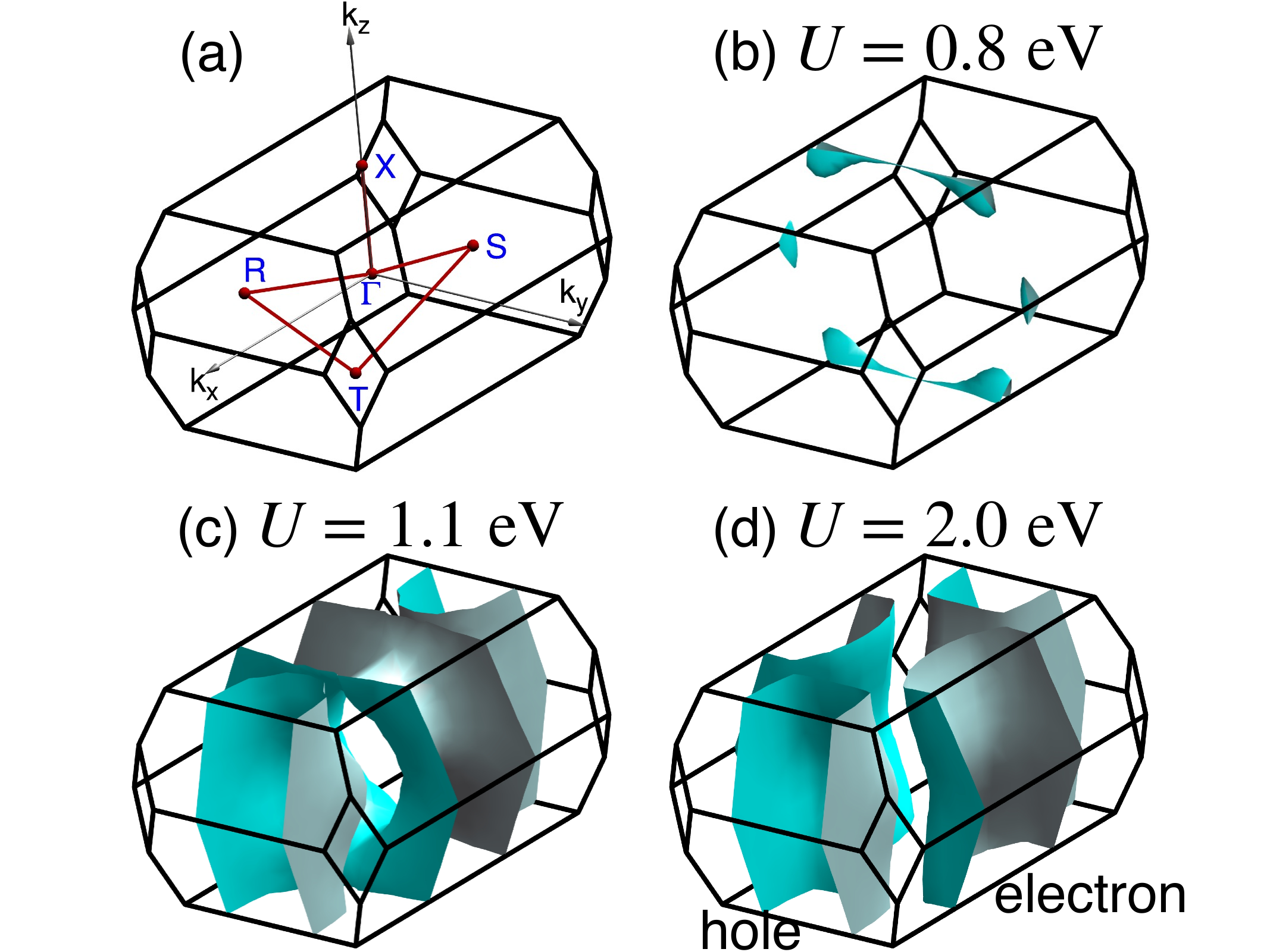}
  \caption{
    (Color online)
    (a) First Brillouin zone and $k$-path of \ute.
    Fermi surfaces calculated by the GGA+$U$ method with (b) 
    $U = 0.8\ev$, (c) 
    $U = 1.1\ev$, and (d) 
    $U = 2.0\ev$.}
  \label{fig:ambientpressure_fs}
\end{figure}

Consistent with previous studies~\cite{Ishizuka2019,Harima2020}, an insulating state is obtained for $U = 0$.
As $U$ increases up to $0.5\ev$, bands start to cross the Fermi level, forming small Fermi surfaces [Fig.~\ref{fig:ambientpressure_fs}(b)].
With $U = 1.1\ev$, a three-dimensional Fermi surface is formed, consisting of a ringlike electron sheet and a cylindrical hole sheet with strong warping [Fig.~\ref{fig:ambientpressure_fs}(c)].
As $U$ is slightly increased to $1.25 \ev$, the electron sheet becomes disconnected at the X point, indicating a topological change from a ringlike to a cylindrical shape. This change is referred to as the $U$-driven Lifshitz transition in this paper.
With further increase in $U$ ($\gtrsim 1.5\ev$), warping of the two Fermi sheets become smaller, and an approximately two-dimensional electronic structure is obtained [Fig.~\ref{fig:ambientpressure_fs}(d)].
This $U$ dependence of electronic structure is qualitatively equivalent to that demonstrated in the previous study~\cite{Ishizuka2019}.
A slight difference is found in the value of $U$ at which the Fermi surface causes the $U$-driven Lifshitz transition.
For $U = 1.5\ev$, our results based on the FPLO method show cylindrical Fermi surfaces, 
while a three-dimensional Fermi surface is obtained in the Wien2k calculation~\cite{Ishizuka2019}. 
The discrepancy is minor and is attributed to differences in the DFT methods rather than in the structural parameters.
The band structures for larger $U > 2\ev$ were calculated in the previous GGA+$U$ and DFT+DMFT calculations~\cite{Ishizuka2019,Xu2019,Miao2020}, and qualitatively the same Fermi surfaces as those for $U=2\ev$ have been obtained.

The dHvA oscillations have been observed in high-quality single crystals of \ute in the high magnetic field region~\cite{Aoki_dHvA2022,Aoki_dHvA2023,Broyles2023,Eaton2024,Weinberger2024}. The results support quasi-two-dimensional Fermi surfaces consistent with the GGA+$U$ calculation for $U = 2.0\ev$, while the presence of a three-dimensional Fermi surface was also discussed~\cite{Broyles2023,Miao2020}. 
In contrast, the bulk measurements in the low magnetic field region show more three-dimensional properties~\cite{Eo2022,Ishihara2023,Suetsugu2024}. One possible interpretation is that the Fermi surfaces change under a magnetic field. However, further verification is needed, and in this paper, we study various values of $U$ without focusing on any particular parameter.

\section{Electronic Structure under Hydrostatic Pressure}

To study electronic structures under hydrostatic pressure, we start by interpolating the pressure dependence of the lattice constants obtained in experimental data~\cite{Honda2023}. For ambient pressure, we adopt the lattice constants from Refs.~\cite{Haga2022,Sakai2022} as in Ref.~\cite{Honda2023}. Despite small deviations, the linear interpolation captures the pressure dependence of the lattice constants (see Appendix~\ref{sec:hydro_lattice_constants}).

\begin{figure}[htb]
  \centering
  \includegraphics[width=\linewidth]{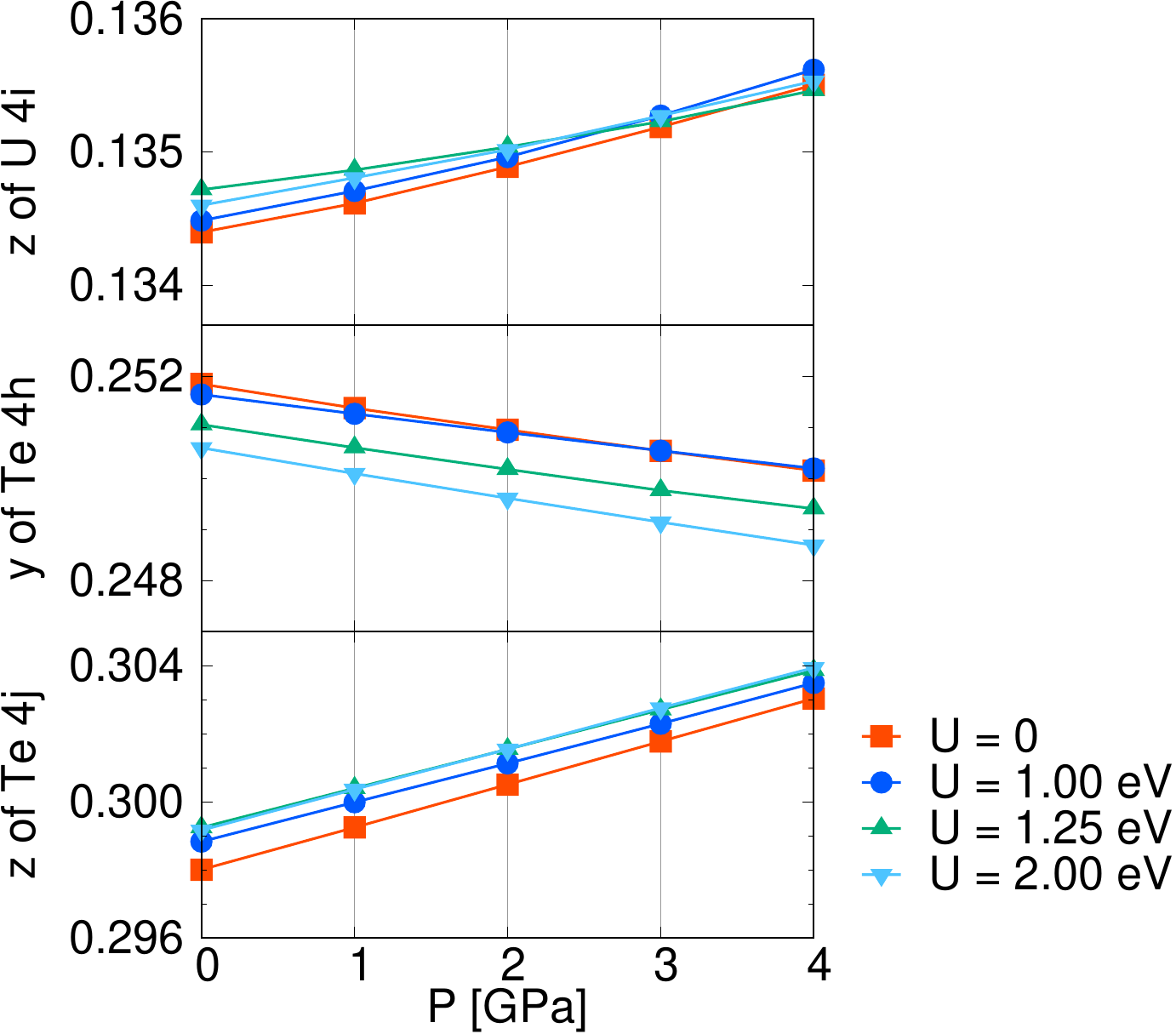}
  \caption{
    (Color online)
    Pressure dependence of fractional coordinates, $z$ of U $4i$, $y$ of Te $4h$, and $z$ of Te $4j$ Wyckoff positions under hydrostatic pressure.
    The ionic relaxation is performed for various $U$ with using the interplated lattice constants in Fig.~\ref{fig:hydrostaticpressure_latticeconstants}.
  }
  \label{fig:hydrostaticpressure_atomiccoordinates}
\end{figure}

Next, we determine the internal atomic positions in the unit cell. There are three Wyckoff positions, U $4i$, Te $4h$ and Te $4j$, whose coordinates are known only at ambient pressure. To determine the atomic coordinates under hydrostatic pressure, we perform ionic relaxations while keeping the lattice constants fixed to the interpolated values. We use the GGA exchange correlation functional~\cite{Perdew1996} and the projector augmented-wave basis as implemented in VASP~\cite{Kresse1996_vasp,Kresse1999_vasp}. We adopt a plane-wave cutoff energy of 600~eV for Kohn-Sham orbitals including spin-orbit coupling and a $12\times12\times12$ $\bm{k}$-mesh. The electronic self-consistent calculations are considered converged when the total energy difference between two consecutive steps is less than $10^{-8}\ev$. We perform ionic relaxation until the Hellmann-Feynman force becomes less than $10^{-4}\ev/\AA$. We execute the relaxation with several values of $U$ from 0 to 2~eV within GGA$+U$ with the Liechtenstein scheme~\cite{Liechtenstein1995_dftu}, which is consistent with the GGA+$U$ method implemented in FPLO.
The obtained atomic coordinates are plotted in Fig.~\ref{fig:hydrostaticpressure_atomiccoordinates}. Changes in atomic coordinates from 0~GPa to 4~GPa are from 0.1\% to 0.5\% of the lattice constants,
indicating that the local structure is uniformly compressed under pressure (see Appendix~\ref{sec:bond_length} for the pressure dependence of characteristic bond lengths).

\begin{figure}[htb]
  \centering
  \includegraphics[width=0.7\linewidth]{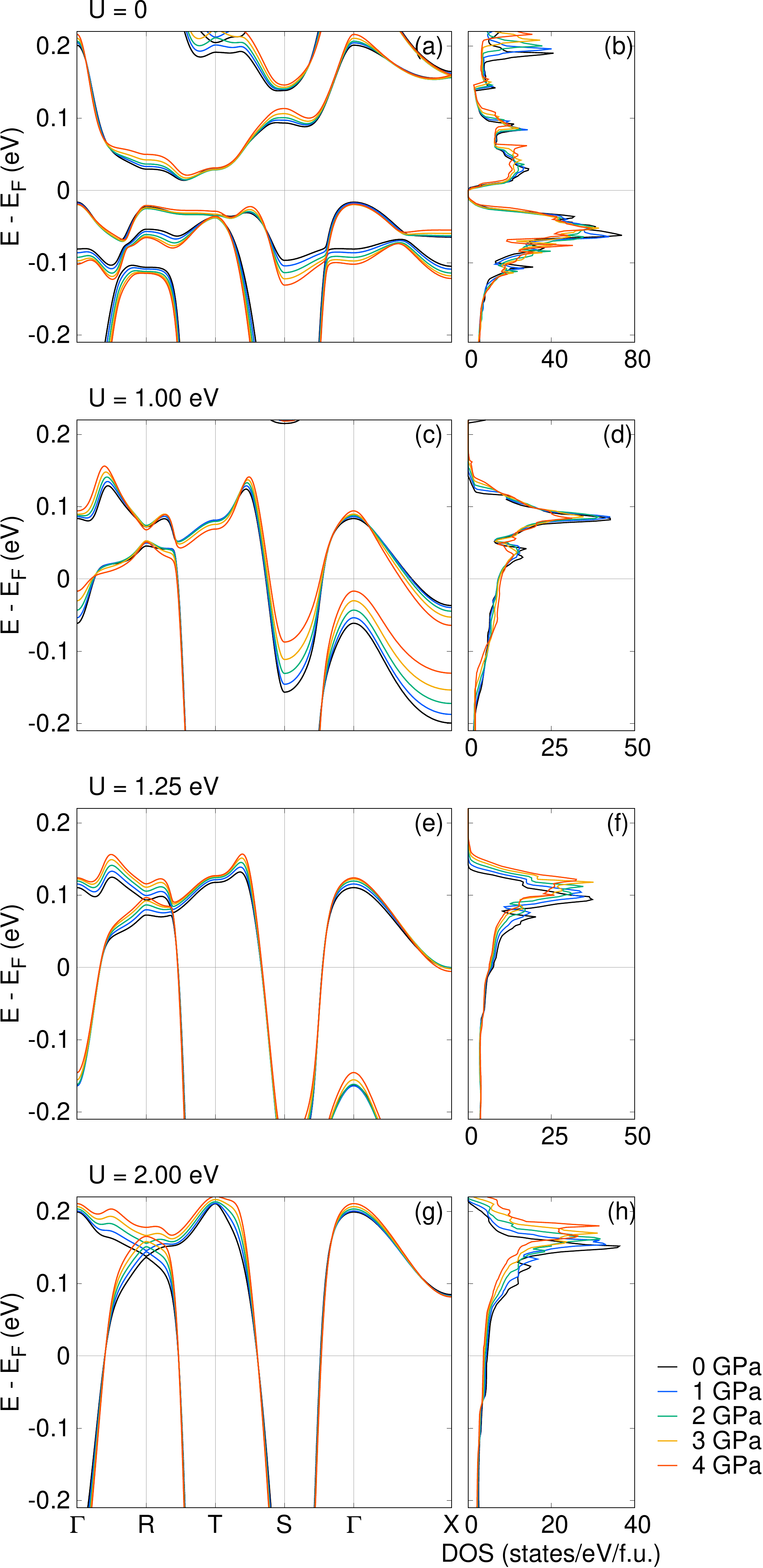}
  \caption{
    (Color online)
    Band structures (left) and DOS (right) under hydrostatic pressure
    obtained by the relativistic GGA+$U$ method with (a, b) $U = 0$, (c, d) $U = 1.0\ev$,
    (e, f) $U = 1.25\ev$, and (g, h) $U = 2.0\ev$.
  }
  \label{fig:hydrostaticpressure_w_relaxation_banddos}
\end{figure}

We calculate electronic structures under hydrostatic pressure
with the interpolated lattice constants
and the relaxed atomic coordinates.
We use the same setting as in the calculations at ambient pressure.
With $U = 0$, which corresponds to GGA,
a small gap opens at the Fermi level, and a insulating state is obtained at ambient pressure [Figs.~\ref{fig:hydrostaticpressure_w_relaxation_banddos}(a,b)].
The band dispersion around the Fermi level does not show significant dependence on pressure, and the insulating state is robust against pressure. In other words, small variations in the crystal structure do not cause the insulator-metal transition. These results for $U = 0$ are incompatible with the experiments, implying that the Coulomb interaction is essential for the metallic state of \ute.

\begin{figure}[htb]
  \centering
  \includegraphics[width=0.9\linewidth]{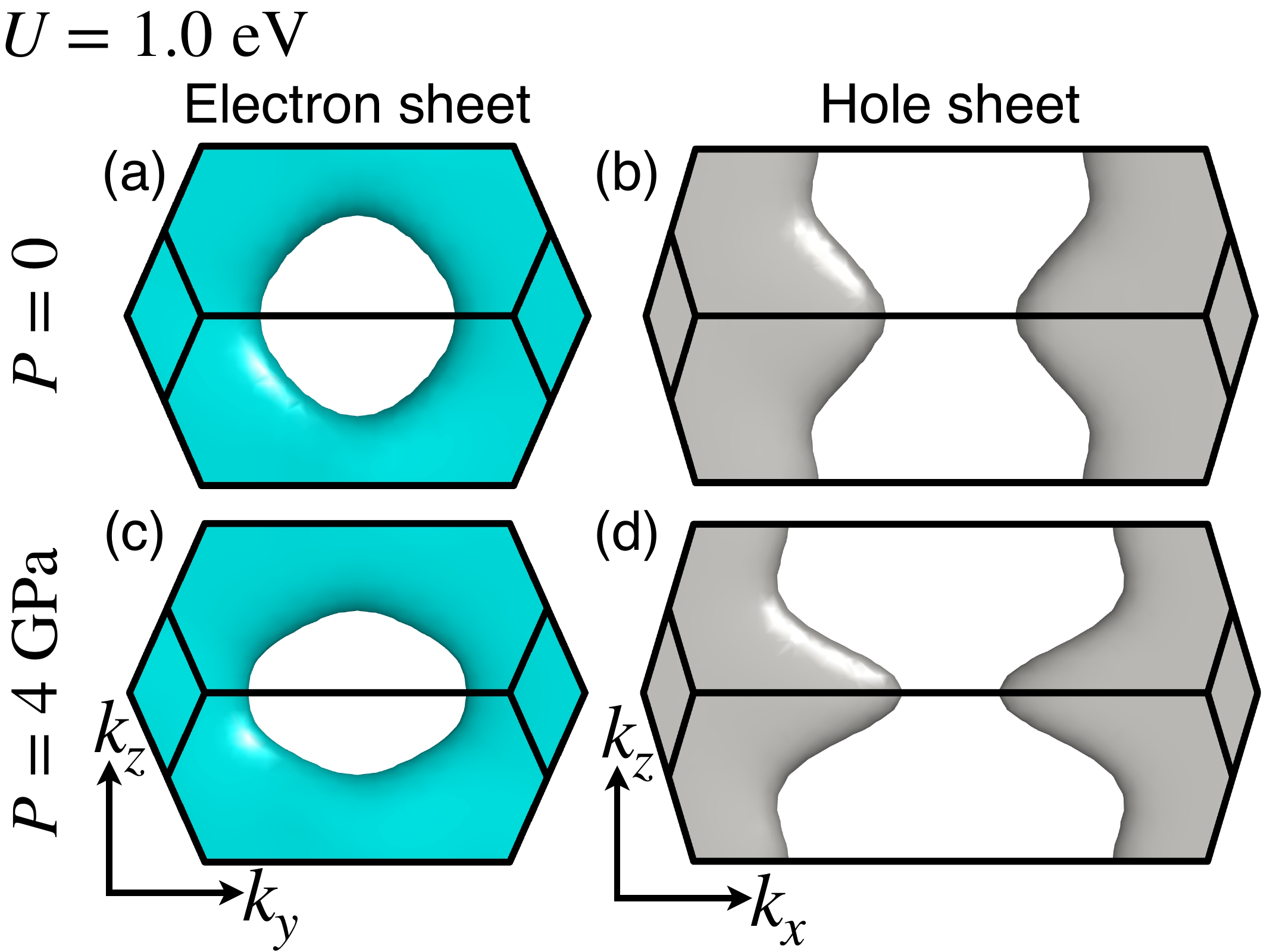}
  \caption{
    (Color online)
    Side views of the electron sheet from the $k_x$ direction (left panels) and side views of the hole sheet from the $k_y$ direction (right panels) at $P=0$ (top panels) and $P=4\gpa$ (bottom panels) obtained by GGA+$U$ calculations for $U = 1.0\ev$.
  }
  \label{fig:hydrostaticpressure_fs_U1p00eV}
\end{figure}

For $U = 1.0\ev$, the bands cross the Fermi level, and a metallic state is obtained [Figs.~\ref{fig:hydrostaticpressure_w_relaxation_banddos}(c,d)]. The Fermi surfaces comprise the three-dimensional ringlike electron sheet and the strongly warped cylindrical hole sheet. 
As increasing pressure, almost all of the local minima and maxima of the valence bands move upward, while the band bottom $\sim -40\mev$ at the X point goes downward. As a result, the band dispersion along the $\Gamma-R$ and $\Gamma-S$ directions decreases while it increases along the $\Gamma-X$ direction. In particular, we see significant band flattening along the $\Gamma-R$ direction, which crosses the hole Fermi sheet. At the same time, density of states (DOS) at the Fermi level increases by 8\% from 0~GPa to 4~GPa [Fig.~\ref{fig:hydrostaticpressure_w_relaxation_banddos}(d)]. This increase would be a consequence of the band flattening. 
On the Fermi level, the shape of the Fermi surfaces changes, and warping of the hole sheet is enhanced under pressure [Fig.~\ref{fig:hydrostaticpressure_fs_U1p00eV}], consistent with the pressure dependence of the band dispersion. 
The effects of pressure on the low-energy band dispersion and the Fermi surface are most significant for this case of $U = 1.0\ev$. 

With $U = 1.25\ev$, the bottom of an electron pocket at the X point is very close to the Fermi level
($E \simeq E_\mathrm{F} - 1\mev$ at $P=0$), and the system is in the vicinity of the $U$-driven Lifshitz transition at ambient pressure.
When pressure is increased, all of the local minima and maxima of valence bands move upward. Although this pressure dependence is qualitatively similar to that observed in the case of $U = 1.0\ev$, the change in the band structure is less significant than for $U = 1.0\ev$. In particular, the band bottom at the X point is insensitive to pressure with this value of $U =1.25\ev$. It decreases by only $3\mev$ when we increase pressure from $P=0$~GPa to 4~GPa, and the electron Fermi surface remains to be three-dimensional.

When we assume a large $U = 2.0\ev$, the electron pocket around the X point disappears [Figs.~\ref{fig:hydrostaticpressure_w_relaxation_banddos}(g,h)], and two-dimensional Fermi surfaces are obtained. Both electron and hole Fermi sheets show a rectangular shape. The band width increases under pressure, and the DOS at the Fermi level decreases. Despite the changes in the Fermi velocity and DOS, the Fermi surfaces almost remain unchanged.

Comparing the results for various $U$, we find that Fermi surfaces do not change so much in most cases by increasing pressure. An exceptional case is $U = 1.0\ev$, where the three-dimensional Fermi surface and the two-dimensional one coexist. The $k_z$-dependence of the energy dispersion is enhanced under pressure.
The U $5f$ occupancy decreases with increasing pressure for all $U$ values, as shown in Appendix~\ref{sec:hydro_occupancy}.

To elucidate the effects of pressure dependence in the lattice structure, we calculate the band structure by fixing the fractional coordinates of the U and Te ions (see Appendix~\ref{sec:hydro_wo_relaxation}). The overall pressure dependence is similar to the results with relaxed atomic coordinates shown in Fig.~\ref{fig:hydrostaticpressure_w_relaxation_banddos}. However, the energy of the band bottom at the X point is more sensitive to pressure than in Fig.~\ref{fig:hydrostaticpressure_w_relaxation_banddos}. In particular, for $U = 1.2\ev$ [Fig.~\ref{fig:hydrostaticpressure_wo_relaxation_banddos}(e)], the band bottom moves upward, and a pressure-induced Lifshitz transition occurs, in which the ringlike electron sheet becomes disconnected at the X point, indicating a topological change from a ringlike to a cylindrical shape. In this sense, the topology of Fermi surfaces is sensitive to atomic positions.
On the other hand, in Fig.~\ref{fig:hydrostaticpressure_wo_relaxation_banddos}(c), the pressure dependence of band dispersion for $U = 1.0\ev$ is much weaker than in Fig.~\ref{fig:hydrostaticpressure_w_relaxation_banddos}(c). This indicates that the significant pressure dependence of the band structure for an intermediate $U = 1.0\ev$ mainly originates from the atomic displacement in the unit cell.

\section{Electronic Structure under Uniaxial Stress}

Next, we discuss electronic structures under uniaxial stress. 
For this purpose, we start by calculating the lattice constants from the elastic moduli estimated by resonant ultrasound spectroscopy (RUS) measurement in Ref.~\cite{Theuss2024_elasticmoduli} (see Appendix~\ref{sec:uniaxial_lattice_constants}).
In this section, we consider the uniaxial stress, $\sigma_{100}$, $\sigma_{010}$, and $\sigma_{001}$, along the $[100]$, $[010]$, and $[001]$ crystallographic directions.
In all cases, the calculated lattice constants at $\sigma=0$ are equivalent to those in TABLE~\ref{tab:ambientpressure_crystalstructure}. The decrease in the lattice constants under uniaxial stress is more significant than under hydrostatic pressure.

\begin{figure}[htb]
  \centering
  \includegraphics[width=0.9\linewidth]{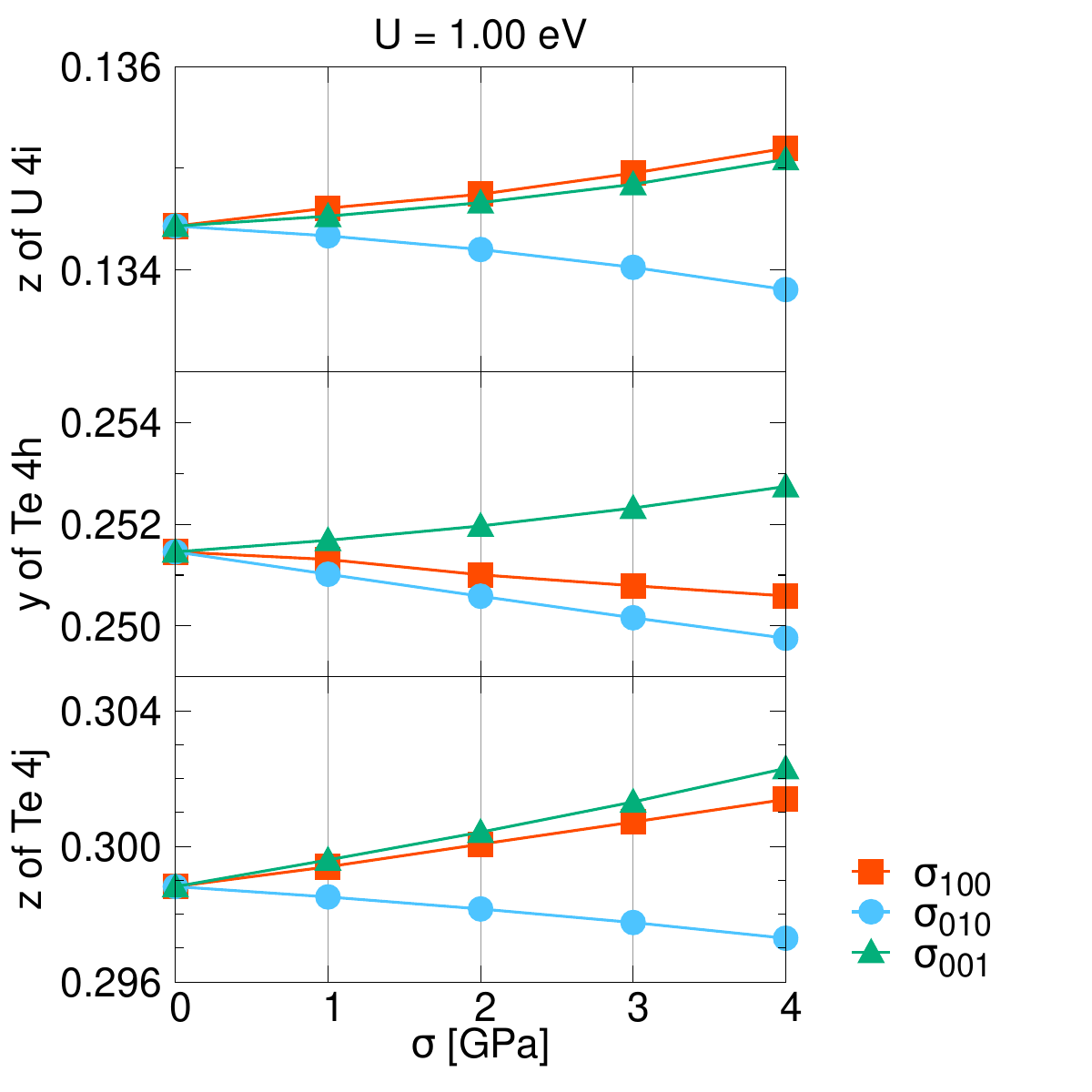}
  \caption{
    (Color online)
    Fractional coordinates  $z$ of U $4i$, $y$ of Te $4h$, and $z$ of Te $4j$ Wyckoff positions under uniaxial stress, $\sigma_{100}$ (red), $\sigma_{010}$ (cyan), and $\sigma_{001}$ (green).
    These results are obtained by the GGA+$U$ method with $U=1.0\ev$.
  }
  \label{fig:uniaxialpressure_atomiccoordinates_U1p00eV}
\end{figure}

We perform ionic relaxation of the atomic coordinates in the same setting as we have done under hydrostatic pressure. Relaxed atomic coordinates are plotted in Fig.~\ref{fig:uniaxialpressure_atomiccoordinates_U1p00eV}. Changes in atomic coordinates from 0~GPa to 4~GPa are from 0.1\% to 0.4\% of the lattice constants.
The changes in local structure reflect the anisotropic lattice distortions under uniaxial stress (see Appendix~\ref{sec:bond_length}).
In the following, we first focus on the GGA+$U$ results for $U = 1.0\ev$.  Although we have also conducted calculations in other cases such as $U =0$ and $2.0\ev$, we do not find significant changes in Fermi surfaces by pressure.

\begin{figure}[htb]
  \centering
  \includegraphics[width=\linewidth]{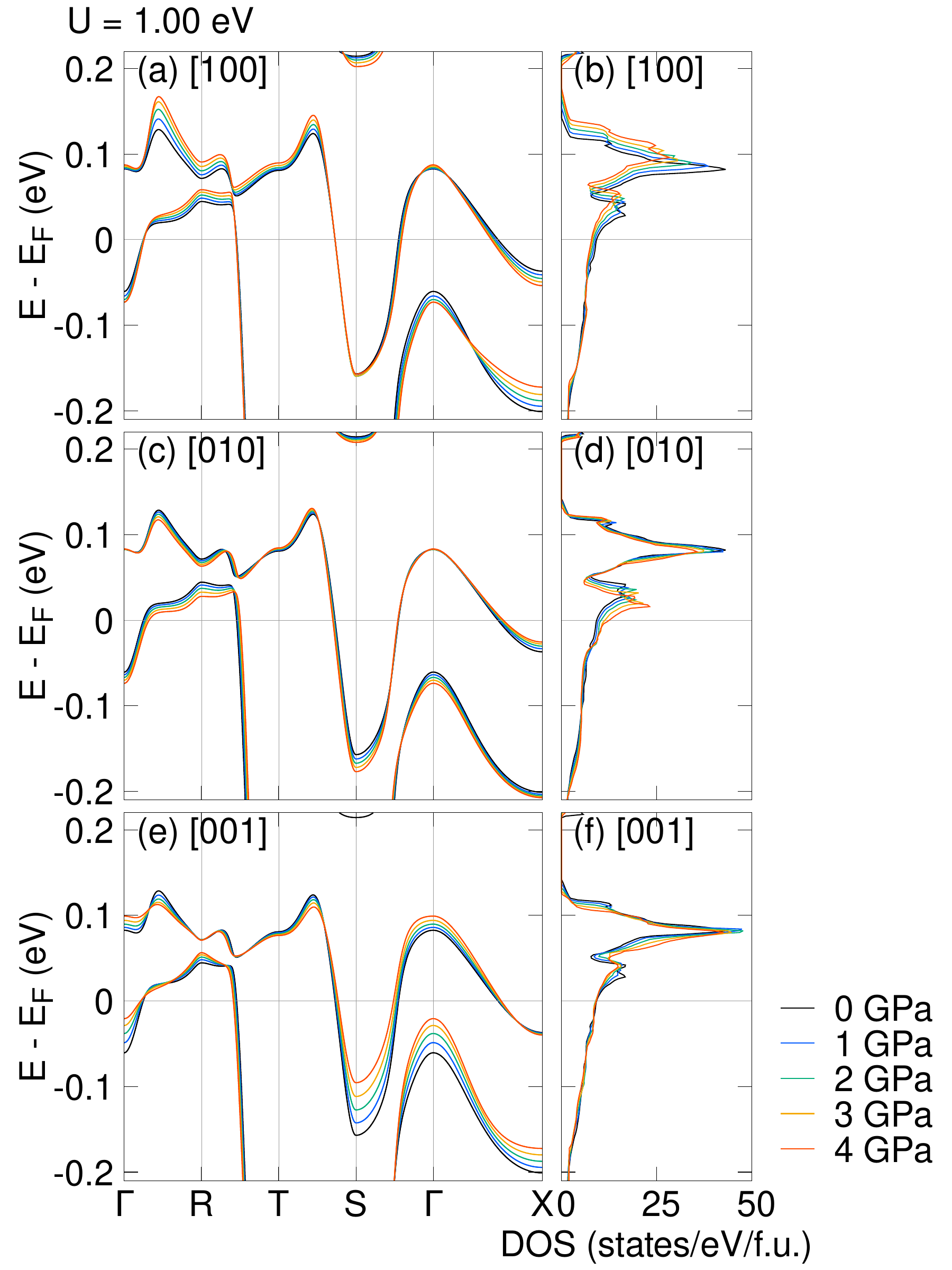}
  \caption{
    (Color online)
    Band structures (left) and DOS (right) under uniaxial stress $\sigma_{100}$ (top), $\sigma_{010}$ (middle), and $\sigma_{001}$ (bottom) calculated with GGA+$U$ for $U = 1.0\ev$ and relaxed atomic positions.
  }
  \label{fig:uniaxialpressure_w_relaxation_banddos_U1p00eV}
\end{figure}

We again perform full relativistic all-electron GGA+$U$ calculations using the FPLO basis on a $12 \times 12 \times 12$ $\bm{k}$-mesh by using the calculated lattice constants and the relaxed atomic coordinates under uniaxial stress. The calculated band structures for $U = 1.0\ev$ are shown in Fig.~\ref{fig:uniaxialpressure_w_relaxation_banddos_U1p00eV}.

First, we discuss the uniaxial stress along the [100] axis. As $\sigma_{100}$ increases, the valence bands become wider and the DOS at the Fermi level decreases under $\sigma_{100}$. The band bottom in the vicinity of the Fermi level at the X point moves downwards and away from the Fermi level. This change inflates the Fermi surface around the X point.

The uniaxial stress along the [010] axis $\sigma_{010}$ leads to a qualitatively different shift in the valence bands. As shown in Fig.~\ref{fig:uniaxialpressure_w_relaxation_banddos_U1p00eV}(c), the valence bands are totally shifted downward.  However, the band bottom at the X point moves upward to conserve the electron density. Interestingly, the dispersion of the unoccupied low-energy band along the $\Gamma-R$ line becomes significantly small, and a nearly flat band appears at $\sigma_{010}=4\gpa$~[Fig.~\ref{fig:uniaxialpressure_w_relaxation_banddos_U1p00eV}(c)].
The flat band at $\sim0.010\ev$ is slightly above the Fermi level.
In Fig.~\ref{fig:uniaxialpressure_w_relaxation_banddos_U1p00eV}(d), we see a corresponding peak in the DOS, which becomes larger and moves close to the Fermi level as increasing $\sigma_{010}$. Due to this flat band just above the Fermi level, the DOS($E_\mathrm{F}$) increases by 36\% from ambient pressure to $\sigma_{010}=4\gpa$.

In contrast to the above two cases, as the uniaxial stress along the [001] axis $\sigma_{001}$ increases, almost all the local minima and maxima of the valence bands move upward~[Fig.~\ref{fig:uniaxialpressure_w_relaxation_banddos_U1p00eV}(e)],
while the band maxima at $\sim 0.02\ev$ between the $\Gamma$ and R points and the band bottom at $\sim -0.04\ev$ at the X point slightly move downward.
As a result, the band dispersion along the $\Gamma-R$ direction decreases while it increases along the $\Gamma-X$ direction.
This means that the band structure shows larger $k_z$-dependence under the uniaxial stress. This feature is similar to what we observed under hydrostatic pressure. Combined with the flattening of the band due to uniaxial stress along the [010] axis, the effects of [001] uniaxial stress qualitatively reproduce the evolution of the band structure under hydrostatic pressure. Therefore, compression along the [001] axis and the [010] axis is expected to play an essential role in the deformation of Fermi surfaces.

To further examine the possibility of uniaxial-stress-induced Lifshitz transitions, additional calculations were performed for $U = 1.25\ev$, as summarized in Appendix~\ref{sec:uniaxial_u1p25ev}. The overall pressure evolution of the electronic structure resembles that for $U = 1\ev$, but with a smaller enhancement of the DOS and the emergence of a $\sigma_{100}$-induced Lifshitz transition. As $\sigma_{100}$ increases, the electron sheet connects at the X point, representing a topological change from a cylindrical to a ringlike shape [Fig.~\ref{fig:uniaxialpressure_fs_ele_U5f_U1p25eV}].

\section{Discussion}

We performed GGA+$U$ calculations for high-quality single crystals of \ute synthesized in 2021~\cite{Sakai2022}. 
Employing ionic relaxation, we determined the crystal structure under hydrostatic pressure and uniaxial stress and calculated the electronic band structure for various Coulomb interactions $U$ and pressures $P$. 
The results at ambient pressure are consistent with the previous studies~\cite{Ishizuka2019,Harima2020} for the structure parameters of intermediate-quality samples. The insulator-to-metal transition and the $U$-driven Lifshitz transition from a ringlike to a cylindrical electron sheet emerge as $U$ increases~\cite{Ishizuka2019}. 
The consistency is reasonable since the lattice constants and the atomic coordinates are not significantly different between the crystals.

Under hydrostatic pressure, the low-energy band structure and Fermi surfaces do not show significant changes in most cases when we fix $U$. 
The band width increases and the DOS decreases with pressure, as would normally be expected. 
However, we see a significant pressure dependence when we set $U =1.0\ev$, where the low-energy band structure, containing substantial $f$-electron components, more significantly depends on $k_z$ under hydrostatic pressure than at ambient pressure. 
The enhancement of three-dimensionality under hydrostatic pressure was reported in a recent experiment~\cite{weinberger2024pressureenhancedfelectronorbitalweighting}, although the Fermi surfaces are cylindrical and the experimental results are obtained in a high magnetic field.

The possibility of three-dimensional topological superconductivity in \ute, which can host Majorana fermions, can be determined by the topology of the electron sheet~\cite{Ishizuka2019,Kreisel2024}.
Majorana surface states,~\cite{Ishizuka2019} whereas this is not the case for the cylindrical electron sheet.
Therefore, changes in the band topology around the X point crucially affect whether the superconducting state is topologically nontrivial or not.
In our study, when ionic relaxation is performed, the pressure-induced topological transition is unlikely to occur. This is because the band bottom around the X point remains almost unchanged under pressure, even when the system is close to the topological transition ($U \simeq 1.25\ev$). Consequently, the topology of the Fermi surface, and hence that of superconductivity, does not change with pressure in this case. In contrast, a pressure-induced topological transition occurs when the fractional atomic coordinates are fixed. In this case, the electron sheet changes from a ringlike to a cylindrical shape, and thus the topological superconducting state may change from a nontrivial to a trivial one as pressure increases.  

Intensive studies of \ute have clarified the rich phase diagram under pressure~\cite{Braithwaite2019,Lin2020,Thomas2020,Aoki2020,Knebel2020,Ran2020,Valiska2021,Aoki2021,knafo2025incommensurateantiferromagnetismute2pressure,Kinjo2023}: Multiple superconducting phases appear at $0 \leq P \leq 1.8\gpa$ and antiferromagnetic order occurs at $1.8\gpa < P$. 
In light of the phase diagram showing various quantum phases, it is expected that the underlying electronic structure changes under pressure.
Let us propose three possibilities of pressure evolution of electronic structure based on the GGA+$U$ calculations.
First, the results for an intermediate Coulomb interaction $U \simeq 1.0\ev$ may be relevant for \ute. 
In this case, the band structure becomes more isotropic under pressure in the sense that the dispersion around the three-dimensional electron Fermi surface and the two-dimensional hole Fermi surface becomes more $k_z$-dependent. 
At the same time, the low-energy band around the two-dimensional part of the Fermi surfaces, such as along the $\Gamma-R$ line, becomes flat, which increases DOS around the Fermi level even when the band width increases. This increase in DOS may favor antiferromagnetic order.
Second, a pressure-induced Lifshitz transition may occur. This scenario is not supported by our results with relaxed ionic positions and fixed $U$. However, experimental data for atomic positions under pressure are desired for better prediction, because the Lifshitz transition at the $X$ point is sensitive to atomic positions. If the Lifshitz transition occurs, not only the topological invariants but also the magnetic and superconducting correlations naturally change with pressure, because the electronic states around the $X$ point have substantial weight of $f$ electrons. 
Third, the value of $U$ should depend on the pressure. As we see in the electronic structures at ambient pressure, larger $U$ makes the Fermi surfaces two-dimensional and their shape rectangular. Therefore, assuming the increase in $U$ by pressure, one would expect the two-dimensional rectangular Fermi surfaces at high pressure.
Such Fermi surfaces favor antiferromagnetic order as a result of the nesting property.
In fact, the assumption of increasing $U$ is compatible with the superconducting phase diagram under pressure and magnetic field~\cite{Vasina2024}. It is reported that the high-pressure and high-field superconducting phases are smoothly connected without phase transition. We can expect that the Coulomb interactions of $f$-electrons effectively increase under the magnetic field because the localized nature is normally enhanced. When we consider that similar electronic structures lead to similar superconducting states, the increase of $U$ under pressure is expected.
In the first and third scenarios, enhanced antiferromagnetic correlations are expected, which is not inconsistent with the experimental observations under hydrostatic pressure, although the evaluation of magnetic instability is desired.
Superconducting phases are expected to change with the pressure evolution of electronic structures and magnetic correlations, which is a future topic of our study. 

Under uniaxial stress, we have obtained the atomic coordinates by ionic relaxation and calculated the electronic structures with various $U$.
The overall pressure evolution of band structure in the uniaxial stress along the [010] direction stands in contrast to that along the [100] and [001] directions when we adopt $U= 1.0\ev$. 
Such anisotropic properties in the electronic structure may be related to the anisotropic dependence on the uniaxial stress~\cite{Theuss2024}, although further calculations are required for a quantitative discussion.

\section*{Acknowledgements}

We acknowledge useful discussions with D.~Aoki, D.~Braithwaite,  R.~Hakuno, J.~Ishizuka, H.~O.~Jeschke, K.~Kuroki, and J.~Tei.
The computation in this paper was done
using the facilities of the Supercomputer Center,
the Institute for Solid State Physics,
the University of Tokyo.
This work was supported by JSPS KAKENHI Grant Numbers JP22H01181, JP23K19032, JP22H04933, JP23K22452, JP23K17353, JP24K21530, JP24H00007.

\appendix
\renewcommand{\thefigure}{\thesection.\arabic{figure}}
\renewcommand{\thetable}{\thesection.\arabic{table}}

\section{Interpolation of Lattice Constants}
\label{sec:hydro_lattice_constants}
\setcounter{figure}{0}
\setcounter{table}{0}

\begin{figure}[htb]
  \centering
  \includegraphics[width=0.8\linewidth]{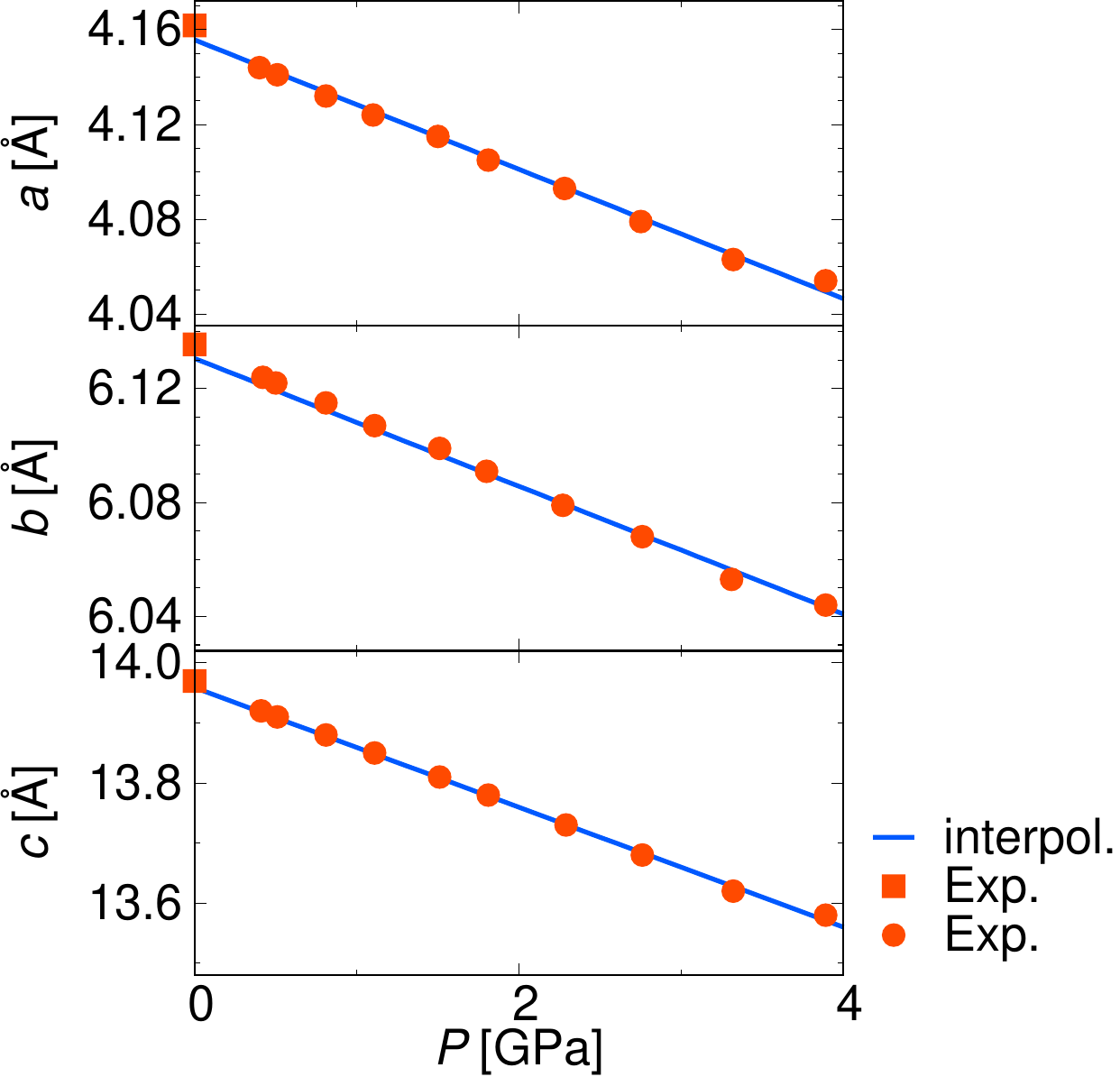}
  \caption{
    (Color online)
    Hydrostatic pressure dependence of the lattice constants, $a$ (top), $b$ (middle), and $c$ (bottom).
    Red squares are experimental values at ambient pressure~\cite{Sakai2022},
    and red circles show experimental data under hydrostatic pressure~\cite{Honda2023}.
    Blue lines show the results of linear interpolation.
  }
  \label{fig:hydrostaticpressure_latticeconstants}
\end{figure}

Lattice constants of \ute at ambient pressure~\cite{Sakai2022} and under pressure~\cite{Honda2023} have been obtained by experiments, 
and we interpolate them.
We plot the experimental data of high-quality single crystals and the interpolating lines of lattice constants in Fig.~\ref{fig:hydrostaticpressure_latticeconstants}. 
The linear interpolation adequately fits the pressure dependence of the experimentally measured lattice constants.

\section{Pressure Dependence of Bond Lengths}
\label{sec:bond_length}
\setcounter{figure}{0}
\setcounter{table}{0}

\begin{figure*}[htb]
  \centering
  \includegraphics[width=1.0\linewidth]{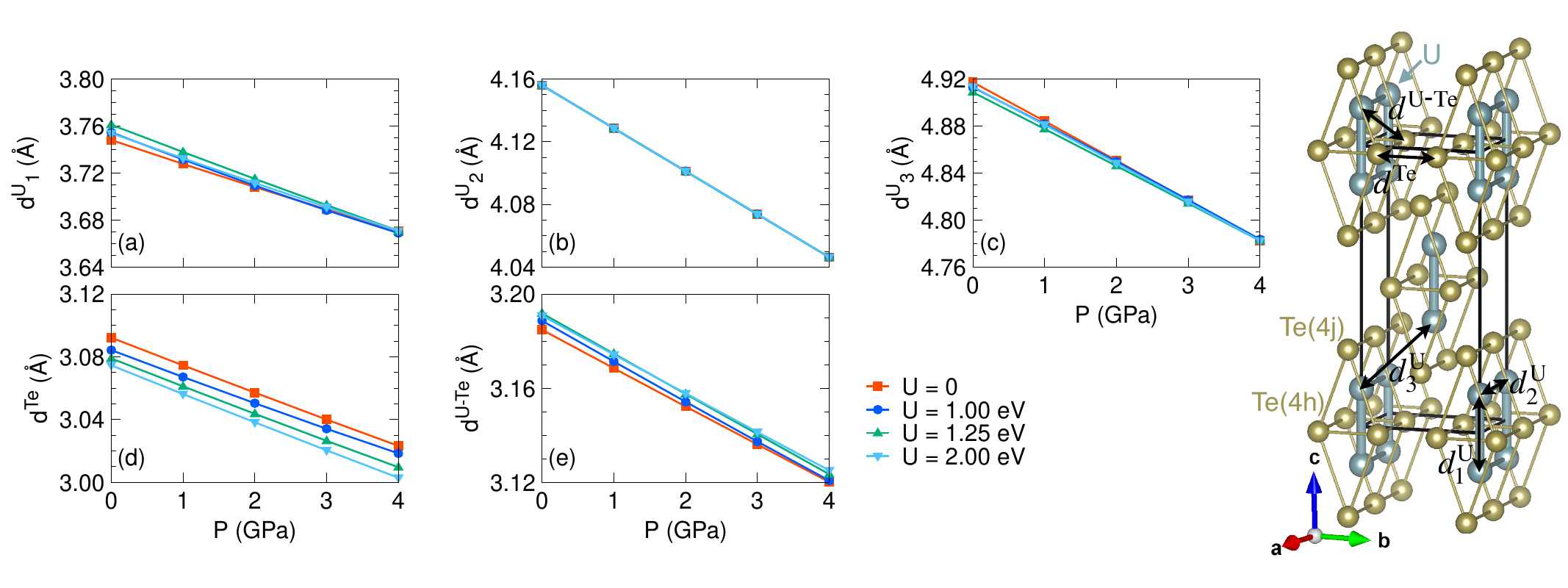}
  \caption{
    (Color online)
    Pressure dependence of bond lengths between U atoms (a-c), between Te atoms (d) and between U and Te atoms (e) calculated for $U = 0$ (red), $U = 1\ev$ (blue), $U = 1.25\ev$ (green) and $U = 2\ev$ (cyan) under hydrostatic pressure. See the right figure for the definitions of the bonds.
  }
  \label{fig:hydrostaticpressure_localgeometry}
\end{figure*}

\begin{figure*}[htb]
  \centering
  \includegraphics[width=1.0\linewidth]{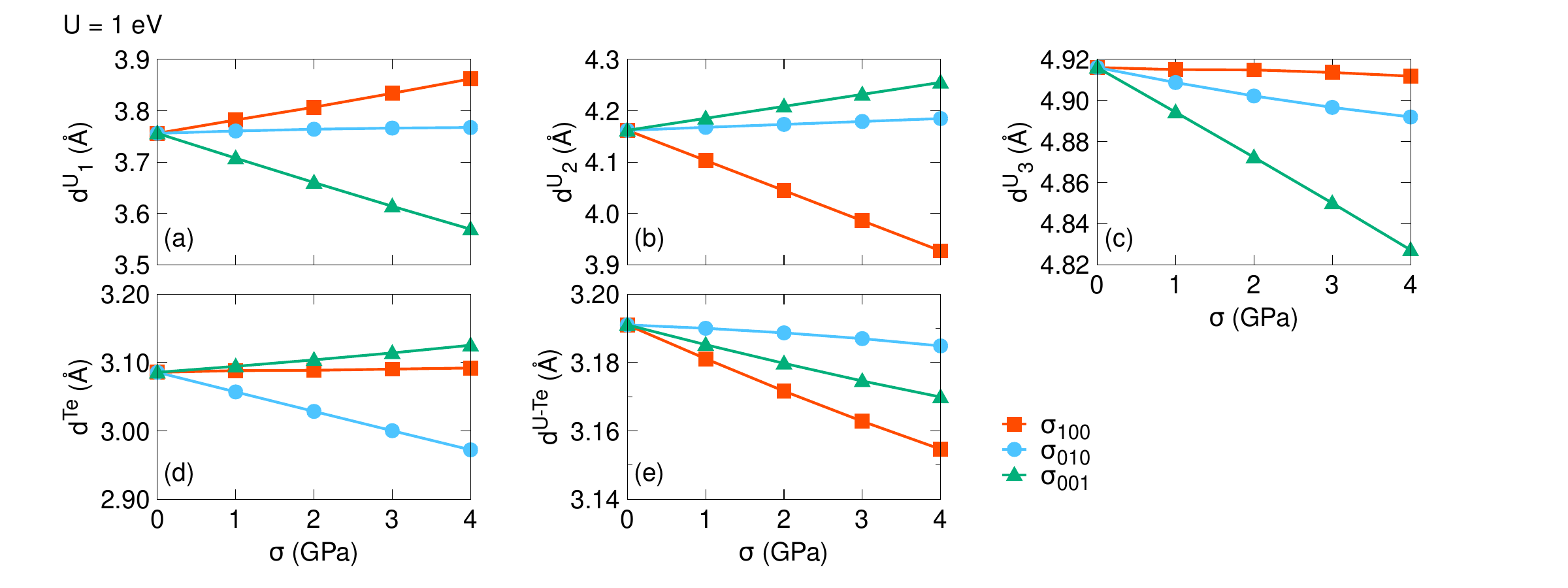}
  \caption{
    (Color online)
    Pressure dependence of bond lengths between U atoms (a-c), between Te atoms (d) and between U and Te atoms (e) calculated for $U = 1\ev$ under uniaxial stress, $\sigma_{100}$ (red), $\sigma_{010}$ (cyan) and $\sigma_{001}$ (green).
  }
  \label{fig:uniaxialpressure_u1p00ev_localgeometry}
\end{figure*}

As a result of ionic relaxation, we show the pressure dependence of bond lengths under hydrostatic pressure (Fig.~\ref{fig:hydrostaticpressure_localgeometry}) and under uniaxial stress (Fig.~\ref{fig:uniaxialpressure_u1p00ev_localgeometry}). The selected bond lengths follow the definition in Ref.~\cite{Ishizuka2021}. $d^\mathrm{U}_1$, $d^\mathrm{U}_2$, $d^\mathrm{U}_3$ represent the nearest-neighbor distances between U atoms, $d^\mathrm{Te}$ represents those between Te atoms, and $d^\mathrm{U\text{-}Te}$ represents those between U and Te atoms.

Under hydrostatic pressure, all bond lengths decrease with increasing pressure at similar rates, indicating a uniform compression of the local structure. The U-U bond lengths ($d^\mathrm{U}_1$, $d^\mathrm{U}_2$, $d^\mathrm{U}_3$) show little sensitivity to the $U$ parameter, whereas the Te-related bond lengths ($d^\mathrm{Te}$ and $d^\mathrm{U\text{-}Te}$) exhibit a more noticeable dependence. This tendency is also seen in Fig.~\ref{fig:hydrostaticpressure_atomiccoordinates}.

Under uniaxial stress, the change rates of the bond lengths are determined by those of the lattice constants. For instance, $d^\mathrm{U}_1$ along the $c$-axis increases under $\sigma_{100}$, increases slowly under $\sigma_{010}$ and decreases under $\sigma_{001}$. This tendency coincides with that of the lattice constants $c$ in the bottom panel of Fig.~\ref{fig:uniaxialpressure_latticeconstants}. In addition, the influences of $\sigma_{010}$ on $d^\text{U}_1$, $d^\text{U}_2$ and $d^\text{U-Te}$ are weaker than those of $\sigma_{100}$ and $\sigma_{001}$. This is related to the $\sigma$-dependence of the lattice constants (Fig.~\ref{fig:uniaxialpressure_latticeconstants}). While $\sigma_{010}$ has the strongest influence on the $b$-axis, as expected, its effect on the other lattice constants is notably weaker compared to the effects of the other uniaxial stress directions. Note that only $d^\text{Te}$ is sensitive to $\sigma_{010}$ compared to the other uniaxial stress directions. This is because the bond is along the [010] direction.

\section{Pressure Dependence of U 5f Occupancy}
\label{sec:hydro_occupancy}
\setcounter{figure}{0}
\setcounter{table}{0}

\begin{figure}[htb]
  \centering
  \includegraphics[width=0.95\linewidth]{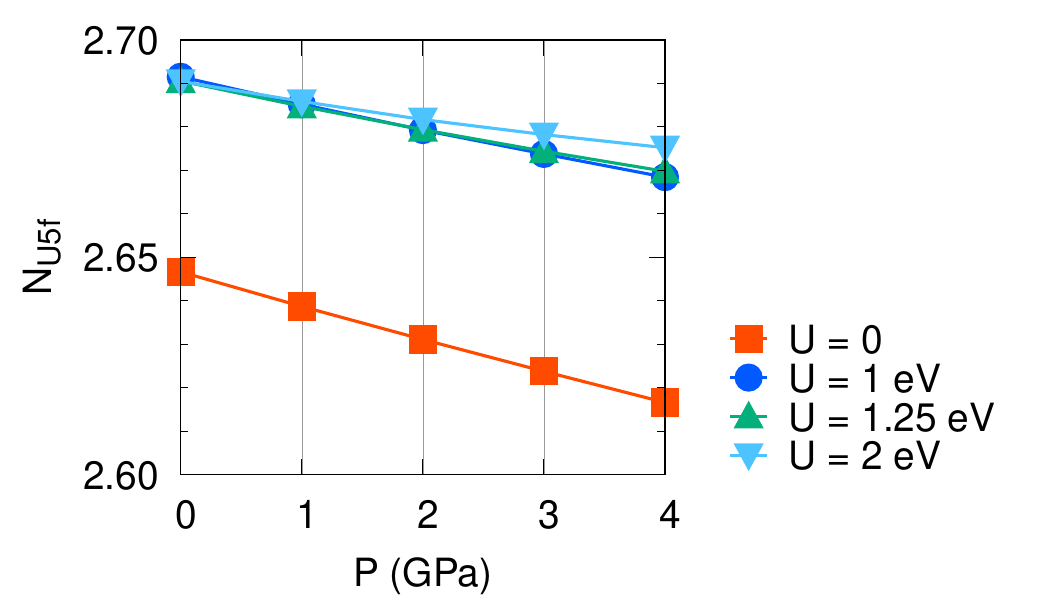}
  \caption{
    (Color online)
    Pressure dependence of U $5f$ occupancy obtained from the GGA+$U$ calculations mentioned in Sec.~III.
  }
  \label{fig:N_net_U5f}
\end{figure}

In Fig.~\ref{fig:N_net_U5f}, we plot pressure dependence of U $5f$ electron occupancy, $N_\mathrm{U5f}$, obtained from the GGA+$U$ calculations mentioned in Sec.~III. At ambient pressure, $N_\mathrm{U5f}$ is 2.65 for $U = 0$ and 2.69 for $U = 1\ev$, $1.25\ev$ and $2\ev$. Under pressure, the $f$-electron occupancy decreases for all values of $U$. Decreases from 0~GPa to 4~GPa are 0.03 for $U = 0$ and 0.02 for $U = 1\ev$, $1.25\ev$ and $2\ev$.

\section{Electronic Structure under Hydrostatic Pressure without Ionic Relaxation}
\label{sec:hydro_wo_relaxation}
\setcounter{figure}{0}
\setcounter{table}{0}

\begin{figure}[htb]
  \centering
  \includegraphics[width=0.7\linewidth]{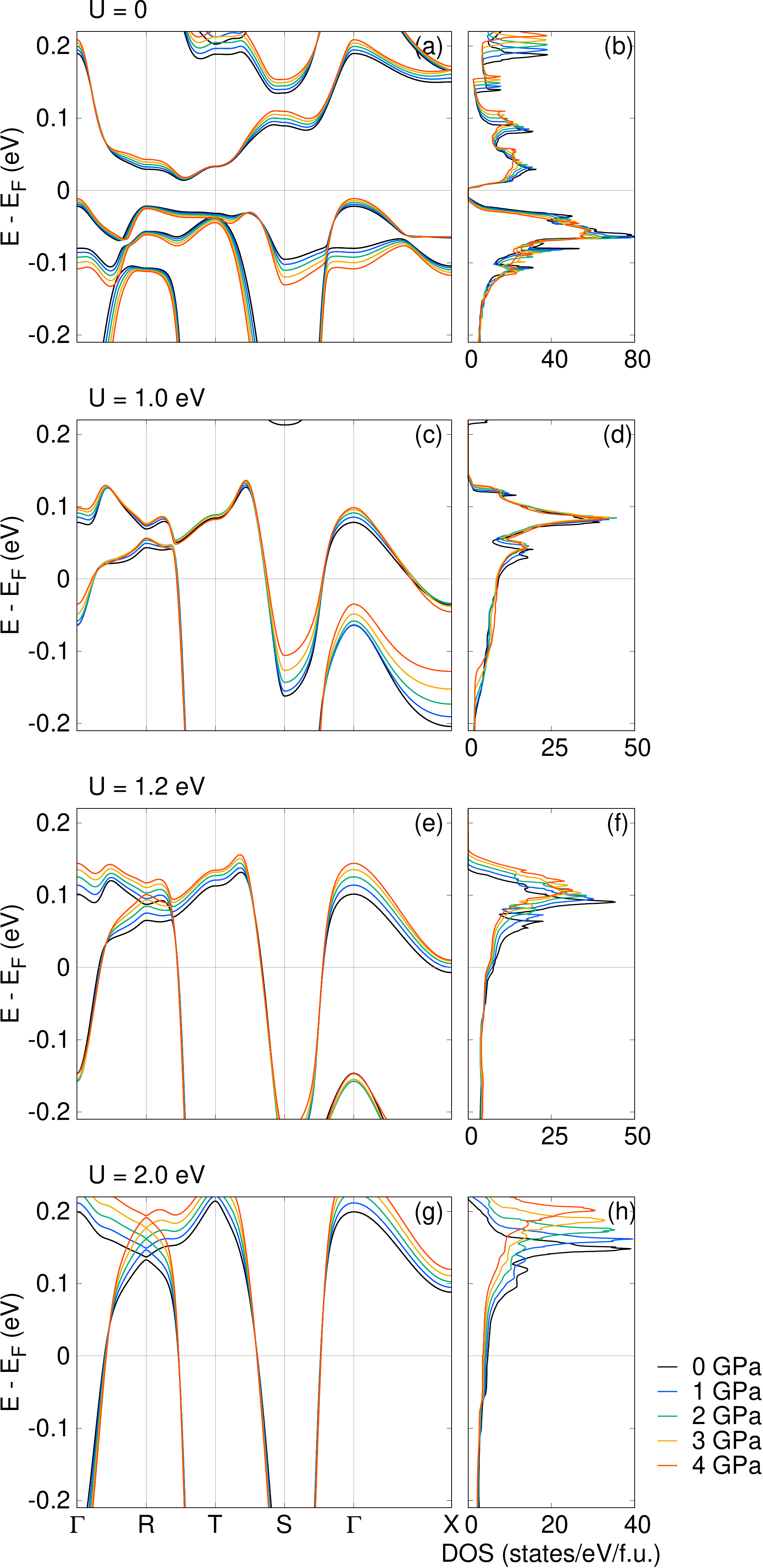}
  \caption{
    (Color online)
    Band structure (left) and DOS (right) under hydrostatic pressure calculated with fixed fractional coordinates of atoms.
    We employ the GGA+$U$ calculation for (a, b) $U = 0$, (c, d) $U = 1.0\ev$, (e, f) $U = 1.2\ev$, and (g, h) $U = 2.0\ev$.
  }
  \label{fig:hydrostaticpressure_wo_relaxation_banddos}
\end{figure}

In the main text, we calculated electronic structures with the relaxed atomic coordinates under pressure.
Here, for comparison, we calculate electronic structures
by fixing the fractional coordinates of atomic positions to be the experimental values at ambient pressure.
In this calculation, the atomic positions vary linearly with the lattice constants.
Figure~\ref{fig:hydrostaticpressure_wo_relaxation_banddos} shows the obtained band structures and DOS.

\section{Lattice Constants under Uniaxial Stress}
\label{sec:uniaxial_lattice_constants}
\setcounter{figure}{0}
\setcounter{table}{0}

\begin{table}[htb]
\centering
\begin{tabular}{cccccc}
\hline
 $\; c_{11}$ & $\; c_{22}$ & $\; c_{33}$ & $\; c_{12}$ & $\; c_{13}$ & $\; c_{23}$ \\
\hline
 \; 89.75 & \; 145.5 & \; 94.95 & \; 26.85 & \; 40.85 & \; 31.65 \\
\hline
\end{tabular}
\caption{
  Elastic moduli in GPa estimated by the RUS measurement~\cite{Theuss2024_elasticmoduli}. They are the average of the data for $T = 4\kelvin$ in Ref.~\cite{Theuss2024_elasticmoduli}. Note that $c_{44}$, $c_{55}$ and $c_{66}$ are not listed since there is no need to consider the shear stress in this study.
}
\label{tab:elasticmoduli}
\end{table}

\begin{figure}[htb]
  \centering
  \includegraphics[width=0.8\linewidth]{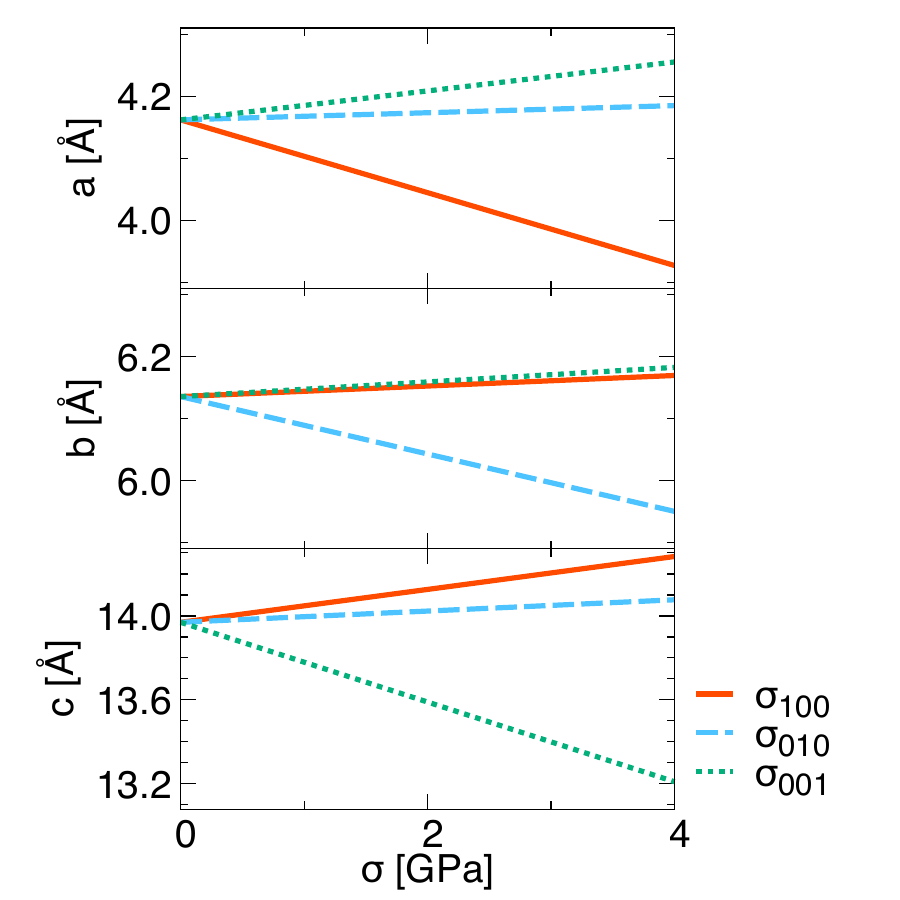}
  \caption{
    (Color online)
    Lattice constants, $a$ (top), $b$ (middle), and $c$ (bottom), under uniaxial stress, $\sigma_{100}$ (solid), $\sigma_{010}$ (dashed), and $\sigma_{001}$ (dotted).
  }
  \label{fig:uniaxialpressure_latticeconstants}
\end{figure}

To evaluate the lattice constants under uniaxial stress, we use the elastic moduli estimated by the RUS measurement~\cite{Theuss2024_elasticmoduli}. In Ref.~\cite{Theuss2024_elasticmoduli}, four variations (two samples and two temperatures) of the elastic moduli are given. We select the moduli at the lower temperature, $T = 4\kelvin$, and take the average of them (TABLE~\ref{tab:elasticmoduli}). With the elastic moduli, we evaluate the lattice constants [Fig.~\ref{fig:uniaxialpressure_latticeconstants}]. The lattice constant parallel to the uniaxial stress naturally decreases, while the other lattice constants increase. From 0~GPa to 4~GPa, $a$ decreases by 5.6~\% under $\sigma_{100}$, $b$  decreases by 3.0~\% under $\sigma_{010}$, and $c$ decreases by 5.4~\% under $\sigma_{001}$.

\section{Electronic Structure under Uniaxial Stress for $U = 1.25\ev$}
\label{sec:uniaxial_u1p25ev}
\setcounter{figure}{0}
\setcounter{table}{0}

\begin{figure}[htb]
  \centering
  \includegraphics[width=0.9\linewidth]{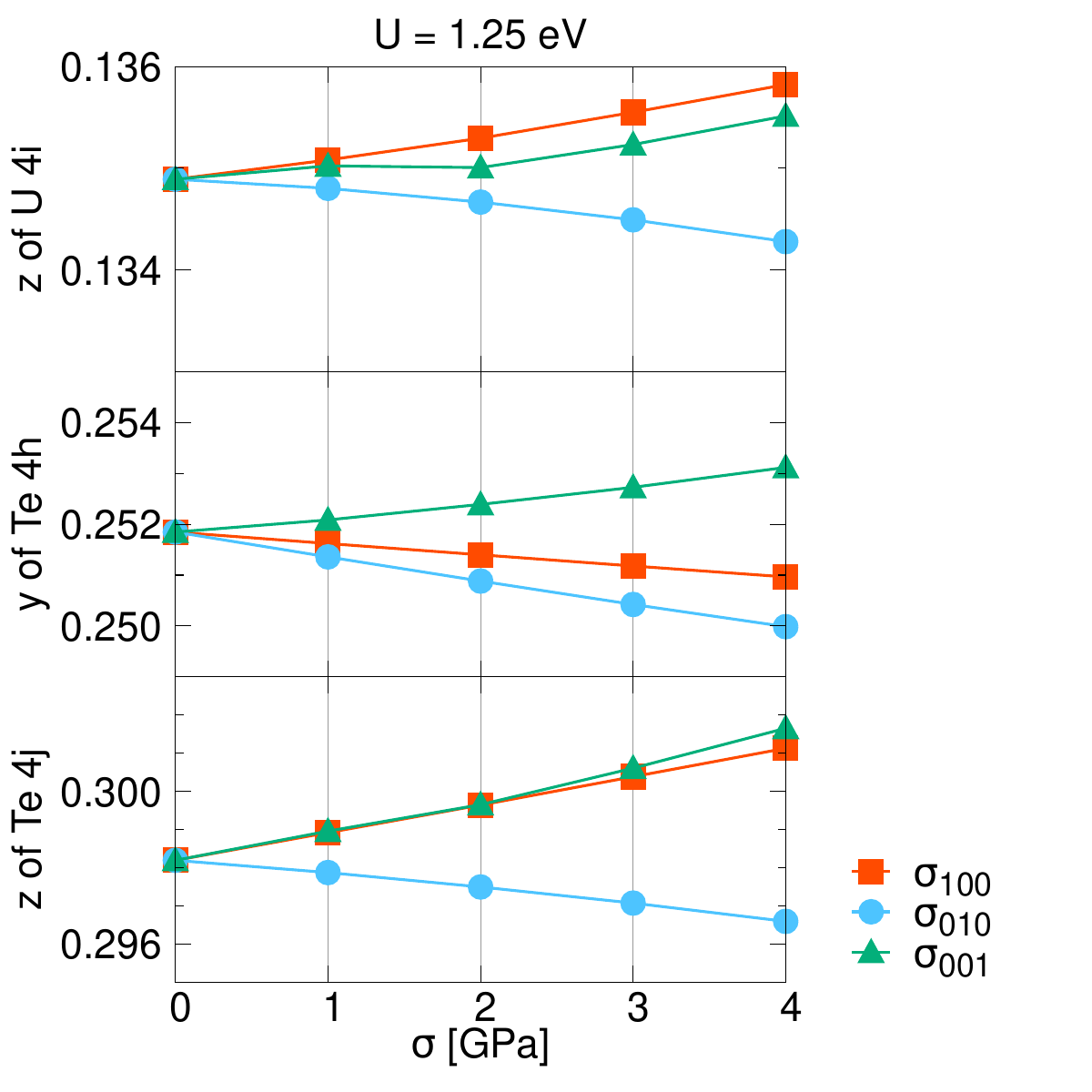}
  \caption{
    (Color online)
    Fractional coordinates  $z$ of U $4i$, $y$ of Te $4h$, and $z$ of Te $4j$ Wyckoff positions under uniaxial stress, $\sigma_{100}$ (red), $\sigma_{010}$ (cyan), and $\sigma_{001}$ (green).
    These results are obtained by the GGA+$U$ method with $U=1.25\ev$.
  }
  \label{fig:uniaxialpressure_atomiccoordinates_U1p25eV}
\end{figure}

\begin{figure}[htb]
  \centering
  \includegraphics[width=\linewidth]{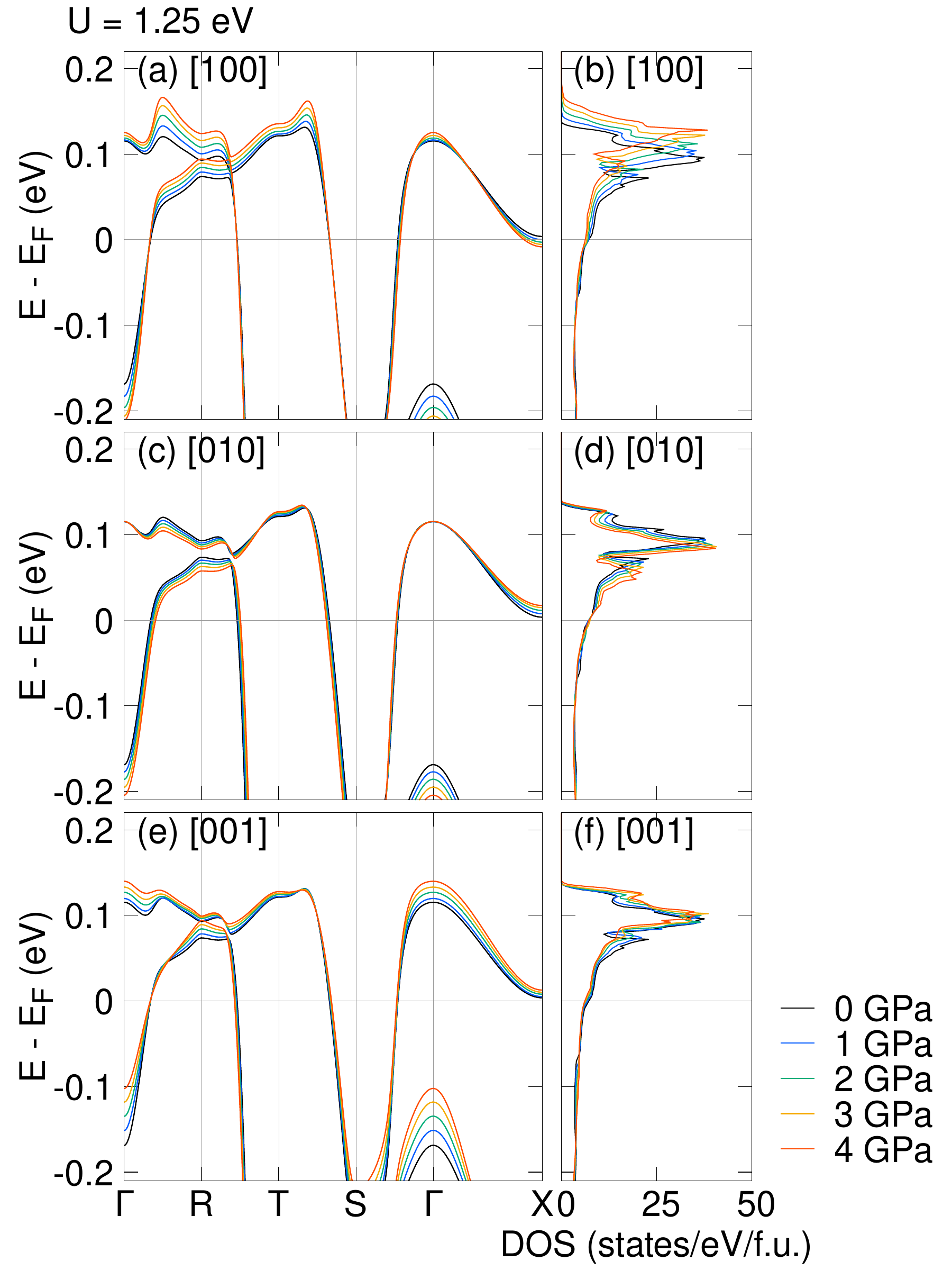}
  \caption{
    (Color online)
    Band structures (left) and DOS (right)
    under uniaxial stress $\sigma_{100}$ (top), $\sigma_{010}$ (middle), and $\sigma_{001}$ (bottom) calculated with 
    GGA+$U$ for $U = 1.25\ev$ and relaxed atomic positions.
  }
  \label{fig:uniaxialpressure_w_relaxation_banddos_U1p25eV}
\end{figure}

We show the results for $U = 1.25\ev$ in Fig.~\ref{fig:uniaxialpressure_atomiccoordinates_U1p25eV} and Fig.~\ref{fig:uniaxialpressure_w_relaxation_banddos_U1p25eV}. Overall pressure evolution of band structures is similar to that for $U=1\ev$. However, we see two different low-energy properties between them. First, for $U=1.25\ev$ an enhancement of the DOS($E_\mathrm{F}$) under $\sigma_{010}$ is notably smaller than that for $U=1\ev$. When we increase the uniaxial stress from $\sigma_{010} = 0$ to $4\gpa$, the DOS($E_\mathrm{F}$) for $U=1.25\ev$ increases by only 0.62\% while that for $U=1\ev$ increases by 36\%. This is because the low-energy unoccupied band on the $\Gamma-R$ path is away from the Fermi level. At $\sigma_{010} = 4\gpa$, the band is around $0.03\ev$ above the Fermi level for $U=1.25\ev$ while that is around $0.01\ev$ for $U=1\ev$. Moreover, the band for $U=1.25\ev$ is not as flat as that for $U=1\ev$. Therefore, the increase of the DOS($E_\mathrm{F}$) for $U=1.25\ev$ is much smaller than that for $U=1\ev$. 

\begin{figure}[htb]
  \centering
  \includegraphics[width=\linewidth]{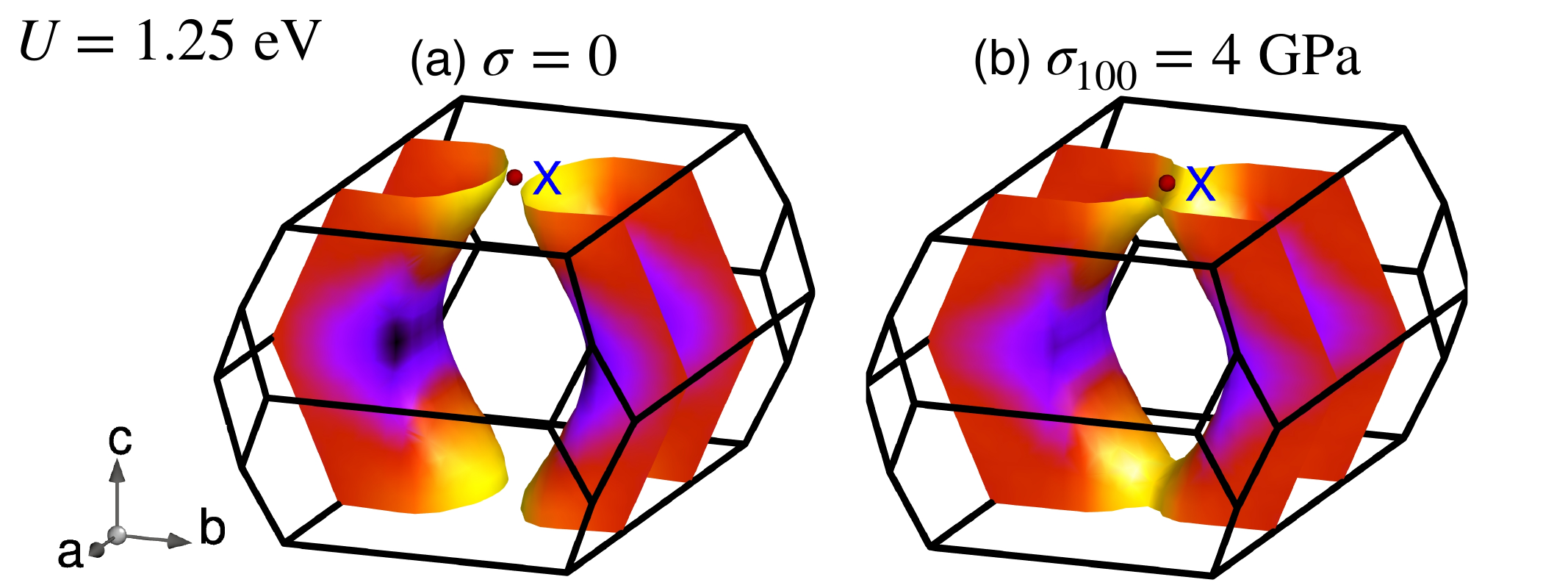}
  \caption{
    (Color online)
    Electron Fermi sheet at (a) ambient pressure and (b) uniaxial stress along the [100] direction $\sigma_{100} = 4.0\gpa$.
    Yellow color indicates a large weight of U $5f$ orbitals as determined by full potential local orbital (FPLO), while violet color indicates a small weight of U $5f$ electrons.
    These results are obtained by the GGA+$U$ calculations for $U=1.25\ev$.
  }
  \label{fig:uniaxialpressure_fs_ele_U5f_U1p25eV}
\end{figure}

Second, a Lifshitz transition occurs under the uniaxial stress $\sigma_{100}$ along the $a$-axis for $U=1.25\ev$, while it does not occur for $U=1\ev$.
Through the Lifshitz transition, the warped cylindrical electron sheet becomes connected at the X point, and the topology of the electron sheet changes from a cylindrical to a ringlike shape. To illustrate the $\sigma_{100}$-induced Lifshitz transition, we plot the electron Fermi sheet at ambient pressure and at $\sigma_{100}=4\gpa$ in Fig.~\ref{fig:uniaxialpressure_fs_ele_U5f_U1p25eV}. Note that change in the DOS$(E_\mathrm{F})$ by the Lifshitz transition is considerably small.

\bibliographystyle{jpsj}
\bibliography{UTe2}

@article{weinberger2024pressureenhancedfelectronorbitalweighting,
  author={T. I. Weinberger and Z. Wu and A. J. Hickey and D. E. Graf and G. Li and P. Wang and R. Zhou and A. Cabala and J. Pu and V. Sechovsky and M. Valiska and G. G. Lonzarich and F. M. Grosche and A. G. Eaton},
  title={Pressure-enhanced $f$-electron orbital weighting in UTe2 mapped by quantum interferometry},
  journal = {arXiv:2403.03946},
  year = {2024},
  url = {https://arxiv.org/abs/2403.03946}
}

@article{Kreisel2024,
  title = {Thermodynamic transitions and topology of spin-triplet superconductivity: Application to ${\mathrm{UTe}}_{2}$},
  author = {R\o{}ising, Henrik S. and Geier, Max and Kreisel, Andreas and Andersen, Brian M.},
  journal = {Phys. Rev. B},
  volume = {109},
  issue = {5},
  pages = {054521},
  numpages = {18},
  year = {2024},
  month = {Feb},
  publisher = {American Physical Society},
  doi = {10.1103/PhysRevB.109.054521},
  url = {https://link.aps.org/doi/10.1103/PhysRevB.109.054521}
}

@article{Weinberger2024,
  title = {Quantum Interference between Quasi-2D Fermi Surface Sheets in ${\mathrm{UTe}}_{2}$},
  author = {Weinberger, T. I. and Wu, Z. and Graf, D. E. and Skourski, Y. and Cabala, A. and Posp\'{\i}\ifmmode \check{s}\else \v{s}\fi{}il, J. and Prokle\ifmmode \check{s}\else \v{s}\fi{}ka, J. and Haidamak, T. and Bastien, G. and Sechovsk\'y, V. and Lonzarich, G. G. and Vali\ifmmode \check{s}\else \v{s}\fi{}ka, M. and Grosche, F. M. and Eaton, A. G.},
  journal = {Phys. Rev. Lett.},
  volume = {132},
  issue = {26},
  pages = {266503},
  numpages = {8},
  year = {2024},
  month = {Jun},
  publisher = {American Physical Society},
  doi = {10.1103/PhysRevLett.132.266503},
  url = {https://link.aps.org/doi/10.1103/PhysRevLett.132.266503}
}

@article{Kinjo2023,
author = {Katsuki Kinjo  and Hiroki Fujibayashi  and Hiroki Matsumura  and Fumiya Hori  and Shunsaku Kitagawa  and Kenji Ishida  and Yo Tokunaga  and Hironori Sakai  and Shinsaku Kambe  and Ai Nakamura  and Yusei Shimizu  and Yoshiya Homma  and Dexin Li  and Fuminori Honda  and Dai Aoki },
title = {Superconducting spin reorientation in spin-triplet multiple superconducting phases of UTe<sub>2</sub>},
journal = {Science Advances},
volume = {9},
number = {30},
pages = {eadg2736},
year = {2023},
doi = {10.1126/sciadv.adg2736},
URL = {https://www.science.org/doi/abs/10.1126/sciadv.adg2736},
}

@article{Ran2020,
  title = "Enhancement and reentrance of spin triplet superconductivity in ${\mathrm{UTe}}_{2}$ under pressure",
  author = "Ran, Sheng and Kim, Hyunsoo and Liu, I-Lin and Saha, Shanta R. and Hayes, Ian and Metz, Tristin and Eo, Yun Suk and Paglione, Johnpierre and Butch, Nicholas P.",
  journal = "Phys. Rev. B",
  volume = {101},
  issue = {14},
  pages = {140503},
  numpages = {6},
  year = {2020},
  month = {Apr},
  publisher = {American Physical Society},
  doi = {10.1103/PhysRevB.101.140503},
  url = {https://link.aps.org/doi/10.1103/PhysRevB.101.140503}
}

@article{Xu2019,
  title = {Quasi-Two-Dimensional Fermi Surfaces and Unitary Spin-Triplet Pairing in the Heavy Fermion Superconductor ${\mathrm{UTe}}_{2}$},
  author = {Xu, Yuanji and Sheng, Yutao and Yang, Yi-feng},
  journal = {Phys. Rev. Lett.},
  volume = {123},
  issue = {21},
  pages = {217002},
  numpages = {6},
  year = {2019},
  month = {Nov},
  publisher = {American Physical Society},
  doi = {10.1103/PhysRevLett.123.217002},
  url = {https://link.aps.org/doi/10.1103/PhysRevLett.123.217002}
}

@article{knafo2025incommensurateantiferromagnetismute2pressure,
  title = {Incommensurate Antiferromagnetism in ${\mathrm{UTe}}_{2}$ under Pressure},
  author = {Knafo, W. and Thebault, T. and Raymond, S. and Manuel, P. and Khalyavin, D. D. and Orlandi, F. and Ressouche, E. and Beauvois, K. and Lapertot, G. and Kaneko, K. and Aoki, D. and Braithwaite, D. and Knebel, G.},
  journal = {Phys. Rev. X},
  volume = {15},
  issue = {2},
  pages = {021075},
  numpages = {16},
  year = {2025},
  month = {May},
  publisher = {American Physical Society},
  doi = {10.1103/PhysRevX.15.021075},
  url = {https://link.aps.org/doi/10.1103/PhysRevX.15.021075}
}

@article{Miao2020,
  title = {Low Energy Band Structure and Symmetries of ${\mathrm{UTe}}_{2}$ from Angle-Resolved Photoemission Spectroscopy},
  author = {Miao, Lin and Liu, Shouzheng and Xu, Yishuai and Kotta, Erica C. and Kang, Chang-Jong and Ran, Sheng and Paglione, Johnpierre and Kotliar, Gabriel and Butch, Nicholas P. and Denlinger, Jonathan D. and Wray, L. Andrew},
  journal = {Phys. Rev. Lett.},
  volume = {124},
  issue = {7},
  pages = {076401},
  numpages = {6},
  year = {2020},
  month = {Feb},
  publisher = {American Physical Society},
  doi = {10.1103/PhysRevLett.124.076401},
  url = {https://link.aps.org/doi/10.1103/PhysRevLett.124.076401}
}

@Article{Ishihara2023,
author={Ishihara, Kota
and Roppongi, Masaki
and Kobayashi, Masayuki
and Imamura, Kumpei
and Mizukami, Yuta
and Sakai, Hironori
and Opletal, Petr
and Tokiwa, Yoshifumi
and Haga, Yoshinori
and Hashimoto, Kenichiro
and Shibauchi, Takasada},
title={Chiral superconductivity in UTe2 probed by anisotropic low-energy excitations},
journal={Nature Communications},
year={2023},
month={May},
day={23},
volume={14},
number={1},
pages={2966},
issn={2041-1723},
doi={10.1038/s41467-023-38688-y},
url={https://doi.org/10.1038/s41467-023-38688-y}
}

@article{Suetsugu2024,
author = {Shota Suetsugu  and Masaki Shimomura  and Masashi Kamimura  and Tomoya Asaba  and Hiroto Asaeda  and Yuki Kosuge  and Yuki Sekino  and Shun Ikemori  and Yuichi Kasahara  and Yuhki Kohsaka  and Minhyea Lee  and Youichi Yanase  and Hironori Sakai  and Petr Opletal  and Yoshifumi Tokiwa  and Yoshinori Haga  and Yuji Matsuda },
title = {Fully gapped pairing state in spin-triplet superconductor UTe<sub>2</sub>},
journal = {Science Advances},
volume = {10},
number = {6},
pages = {eadk3772},
year = {2024},
doi = {10.1126/sciadv.adk3772},
URL = {https://www.science.org/doi/abs/10.1126/sciadv.adk3772},
}

@article{Eo2022,
  title = {$c$-axis transport in ${\mathrm{UTe}}_{2}$: Evidence of three-dimensional conductivity component},
  author = {Eo, Yun Suk and Liu, Shouzheng and Saha, Shanta R. and Kim, Hyunsoo and Ran, Sheng and Horn, Jarryd A. and Hodovanets, Halyna and Collini, John and Metz, Tristin and Fuhrman, Wesley T. and Nevidomskyy, Andriy H. and Denlinger, Jonathan D. and Butch, Nicholas P. and Fuhrer, Michael S. and Wray, L. Andrew and Paglione, Johnpierre},
  journal = {Phys. Rev. B},
  volume = {106},
  issue = {6},
  pages = {L060505},
  numpages = {7},
  year = {2022},
  month = {Aug},
  publisher = {American Physical Society},
  doi = {10.1103/PhysRevB.106.L060505},
  url = {https://link.aps.org/doi/10.1103/PhysRevB.106.L060505}
}

@Article{Eaton2024,
author={Eaton, A. G.
and Weinberger, T. I.
and Popiel, N. J. M.
and Wu, Z.
and Hickey, A. J.
and Cabala, A.
and Posp{\'i}{\v{s}}il, J.
and Prokle{\v{s}}ka, J.
and Haidamak, T.
and Bastien, G.
and Opletal, P.
and Sakai, H.
and Haga, Y.
and Nowell, R.
and Benjamin, S. M.
and Sechovsk{\'y}, V.
and Lonzarich, G. G.
and Grosche, F. M.
and Vali{\v{s}}ka, M.},
title={Quasi-2D Fermi surface in the anomalous superconductor UTe2},
journal={Nature Communications},
year={2024},
month={Jan},
day={03},
volume={15},
number={1},
pages={223},
issn={2041-1723},
doi={10.1038/s41467-023-44110-4},
url={https://doi.org/10.1038/s41467-023-44110-4}
}

@article{Broyles2023,
  title = {Revealing a 3D Fermi Surface Pocket and Electron-Hole Tunneling in ${\mathrm{UTe}}_{2}$ with Quantum Oscillations},
  author = {Broyles, Christopher and Rehfuss, Zack and Siddiquee, Hasan and Zhu, Jiahui Althena and Zheng, Kaiwen and Nikolo, Martin and Graf, David and Singleton, John and Ran, Sheng},
  journal = {Phys. Rev. Lett.},
  volume = {131},
  issue = {3},
  pages = {036501},
  numpages = {7},
  year = {2023},
  month = {Jul},
  publisher = {American Physical Society},
  doi = {10.1103/PhysRevLett.131.036501},
  url = {https://link.aps.org/doi/10.1103/PhysRevLett.131.036501}
}

@article{Aoki_dHvA2023,
author = {Aoki ,Dai and Sheikin ,Ilya and McCollam ,Alix and Ishizuka ,Jun and Yanase ,Youichi and Lapertot ,Gerard and Flouquet ,Jacques and Knebel ,Georg},
title = {de Haas–van Alphen Oscillations for the Field Along c-axis in UTe2},
journal = {J. Phys. Soc. Jpn.},
volume = {92},
number = {6},
pages = {065002},
year = {2023},
doi = {10.7566/JPSJ.92.065002},
URL = { 
        https://doi.org/10.7566/JPSJ.92.065002
},
eprint = { 
        https://doi.org/10.7566/JPSJ.92.065002
}
,
}

@article{Aoki_dHvA2022,
author = {Aoki ,Dai and Sakai ,Hironori and Opletal ,Petr and Tokiwa ,Yoshifumi and Ishizuka ,Jun and Yanase ,Youichi and Harima ,Hisatomo and Nakamura ,Ai and Li ,Dexin and Homma ,Yoshiya and Shimizu ,Yusei and Knebel ,Georg and Flouquet ,Jacques and Haga ,Yoshinori},
title = {First Observation of the de Haas–van Alphen Effect and Fermi Surfaces in the Unconventional Superconductor UTe2},
journal = {J. Phys. Soc. Jpn.},
volume = {91},
number = {8},
pages = {083704},
year = {2022},
doi = {10.7566/JPSJ.91.083704},

URL = { 
    
        https://doi.org/10.7566/JPSJ.91.083704
    
    

},
eprint = { 
    
        https://doi.org/10.7566/JPSJ.91.083704
    
    

}
,
}

@article{Kitamura2023,
  title = {Quantum geometry induced anapole superconductivity},
  author = {Kitamura, Taisei and Kanasugi, Shota and Chazono, Michiya and Yanase, Youichi},
  journal = {Phys. Rev. B},
  volume = {107},
  issue = {21},
  pages = {214513},
  numpages = {14},
  year = {2023},
  month = {Jun},
  publisher = {American Physical Society},
  doi = {10.1103/PhysRevB.107.214513},
  url = {https://link.aps.org/doi/10.1103/PhysRevB.107.214513}
}

@article{Chazono2023,
  title = {Piezoelectric effect and diode effect in anapole and monopole superconductors},
  author = {Chazono, Michiya and Kanasugi, Shota and Kitamura, Taisei and Yanase, Youichi},
  journal = {Phys. Rev. B},
  volume = {107},
  issue = {21},
  pages = {214512},
  numpages = {12},
  year = {2023},
  month = {Jun},
  publisher = {American Physical Society},
  doi = {10.1103/PhysRevB.107.214512},
  url = {https://link.aps.org/doi/10.1103/PhysRevB.107.214512}
}

@Article{Kanasugi2022,
author={Kanasugi, Shota
and Yanase, Youichi},
title={Anapole superconductivity from {\$}{\$}{\{}{\{}{\{}{\{}{\{}{\{}{\{}{\backslash}mathcal{\{}PT{\}}{\}}{\}}{\}}{\}}{\}}{\}}{\}}{\$}{\$}-symmetric mixed-parity interband pairing},
journal={Communications Physics},
year={2022},
month={Feb},
day={10},
volume={5},
number={1},
pages={39},
issn={2399-3650},
doi={10.1038/s42005-022-00804-7},
url={https://doi.org/10.1038/s42005-022-00804-7}
}

@article{Tei2024,
  title = {Pairing symmetries of multiple superconducting phases in ${\mathrm{UTe}}_{2}$: Competition between ferromagnetic and antiferromagnetic fluctuations},
  author = {Tei, Jushin and Mizushima, Takeshi and Fujimoto, Satoshi},
  journal = {Phys. Rev. B},
  volume = {109},
  issue = {6},
  pages = {064516},
  numpages = {9},
  year = {2024},
  month = {Feb},
  publisher = {American Physical Society},
  doi = {10.1103/PhysRevB.109.064516},
  url = {https://link.aps.org/doi/10.1103/PhysRevB.109.064516}
}

@article{Knebel2020,
author = {Knebel ,Georg and Kimata ,Motoi and Vali\v{s}ka ,Michal and Honda ,Fuminori and Li ,DeXin and Braithwaite ,Daniel and Lapertot ,G\'{e}rard and Knafo ,William and Pourret ,Alexandre and Sato ,Yoshiki J. and Shimizu ,Yusei and Kihara ,Takumi and Brison ,Jean-Pascal and Flouquet ,Jacques and Aoki ,Dai},
title = {Anisotropy of the Upper Critical Field in the Heavy-Fermion Superconductor UTe2 under Pressure},
journal = {J. Phys. Soc. Jpn.},
volume = {89},
number = {5},
pages = {053707},
year = {2020},
doi = {10.7566/JPSJ.89.053707},

URL = { 
    
        https://doi.org/10.7566/JPSJ.89.053707
    
    

},
eprint = { 
    
        https://doi.org/10.7566/JPSJ.89.053707
    
    

}
,
}

@article{Aoki2019,
author = {Aoki ,Dai and Nakamura ,Ai and Honda ,Fuminori and Li ,DeXin and Homma ,Yoshiya and Shimizu ,Yusei and Sato ,Yoshiki J. and Knebel ,Georg and Brison ,Jean-Pascal and Pourret ,Alexandre and Braithwaite ,Daniel and Lapertot ,Gerard and Niu ,Qun and Vali\v{s}ka ,Michal and Harima ,Hisatomo and Flouquet ,Jacques},
title = {Unconventional Superconductivity in Heavy Fermion UTe2},
journal = {Journal of the Physical Society of Japan},
volume = {88},
number = {4},
pages = {043702},
year = {2019},
doi = {10.7566/JPSJ.88.043702},
URL = { 
        https://doi.org/10.7566/JPSJ.88.043702
},
eprint = { 
        https://doi.org/10.7566/JPSJ.88.043702
},
}

@article{Aoki2020,
author = {Aoki ,Dai and Honda ,Fuminori and Knebel ,Georg and Braithwaite ,Daniel and Nakamura ,Ai and Li ,DeXin and Homma ,Yoshiya and Shimizu ,Yusei and Sato ,Yoshiki J. and Brison ,Jean-Pascal and Flouquet ,Jacques},
title = {Multiple Superconducting Phases and Unusual Enhancement of the Upper Critical Field in UTe2},
journal = {J. Phys. Soc. Jpn.},
volume = {89},
number = {5},
pages = {053705},
year = {2020},
doi = {10.7566/JPSJ.89.053705},

URL = { 
    
        https://doi.org/10.7566/JPSJ.89.053705
    
    

},
eprint = { 
    
        https://doi.org/10.7566/JPSJ.89.053705
    
    

}
,
}

@article{Thomas2020,
author = {S. M. Thomas  and F. B. Santos  and M. H. Christensen  and T. Asaba  and F. Ronning  and J. D. Thompson  and E. D. Bauer  and R. M. Fernandes  and G. Fabbris  and P. F. S. Rosa },
title = {Evidence for a pressure-induced antiferromagnetic quantum critical point in intermediate-valence UTe<sub>2</sub>},
journal = {Science Advances},
volume = {6},
number = {42},
pages = {eabc8709},
year = {2020},
doi = {10.1126/sciadv.abc8709},
URL = {https://www.science.org/doi/abs/10.1126/sciadv.abc8709},
eprint = {https://www.science.org/doi/pdf/10.1126/sciadv.abc8709},
}

@Article{Lin2020,
author={Lin, Wen-Chen
and Campbell, Daniel J.
and Ran, Sheng
and Liu, I-Lin
and Kim, Hyunsoo
and Nevidomskyy, Andriy H.
and Graf, David
and Butch, Nicholas P.
and Paglione, Johnpierre},
title={Tuning magnetic confinement of spin-triplet superconductivity},
journal={npj Quantum Materials},
year={2020},
month={Sep},
day={25},
volume={5},
number={1},
pages={68},
issn={2397-4648},
doi={10.1038/s41535-020-00270-w},
url={https://doi.org/10.1038/s41535-020-00270-w}
}

@article{Sato2016,
author = {Sato ,Masatoshi and Fujimoto ,Satoshi},
title = {Majorana Fermions and Topology in Superconductors},
journal = {J. Phys. Soc. Jpn.},
volume = {85},
number = {7},
pages = {072001},
year = {2016},
doi = {10.7566/JPSJ.85.072001},
URL = {https://doi.org/10.7566/JPSJ.85.072001},
eprint = {https://doi.org/10.7566/JPSJ.85.072001},
}

@article{Aoki_2022review,
doi = {10.1088/1361-648X/ac5863},
url = {https://dx.doi.org/10.1088/1361-648X/ac5863},
year = {2022},
month = {apr},
publisher = {IOP Publishing},
volume = {34},
number = {24},
pages = {243002},
author = {D Aoki and J-P Brison and J Flouquet and K Ishida and G Knebel and Y Tokunaga and Y Yanase},
title = {Unconventional superconductivity in UTe2},
journal = {Journal of Physics: Condensed Matter},
}

@article{Koepernik1999,
  title = {Full-potential nonorthogonal local-orbital minimum-basis band-structure scheme},
  author = {Koepernik, Klaus and Eschrig, Helmut},
  journal = {Phys. Rev. B},
  volume = {59},
  issue = {3},
  pages = {1743--1757},
  numpages = {0},
  year = {1999},
  month = {Jan},
  publisher = {American Physical Society},
  doi = {10.1103/PhysRevB.59.1743},
  url = {https://link.aps.org/doi/10.1103/PhysRevB.59.1743} }

@article{Ylvisaker2009,
  title = {Anisotropy and magnetism in the $\text{LSDA}+\text{U}$ method},
  author = {Ylvisaker, Erik R. and Pickett, Warren E. and Koepernik, Klaus},
  journal = {Phys. Rev. B},
  volume = {79},
  issue = {3},
  pages = {035103},
  numpages = {12},
  year = {2009},
  month = {Jan},
  publisher = {American Physical Society},
  doi = {10.1103/PhysRevB.79.035103},
  url = {https://link.aps.org/doi/10.1103/PhysRevB.79.035103}
}

@article{Perdew1996,
  title = {Generalized Gradient Approximation Made Simple},
  author = {Perdew, John P. and Burke, Kieron and Ernzerhof, Matthias},
  journal = {Phys. Rev. Lett.},
  volume = {77},
  issue = {18},
  pages = {3865--3868},
  numpages = {0},
  year = {1996},
  month = {Oct},
  publisher = {American Physical Society},
  doi = {10.1103/PhysRevLett.77.3865},
  url = {https://link.aps.org/doi/10.1103/PhysRevLett.77.3865}
}

@article{Kresse1996_vasp,
  title = {Efficiency of ab-initio total energy calculations for metals and semiconductors using a plane-wave basis set},
  journal = {Computational Materials Science},
  volume = {6},
  number = {1},
  pages = {15-50},
  year = {1996},
  issn = {0927-0256},
  doi = {https://doi.org/10.1016/0927-0256(96)00008-0},
  url = {https://www.sciencedirect.com/science/article/pii/0927025696000080},
  author = {G. Kresse and J. Furthmüller},
}

@article{Kresse1999_vasp,
  title = {From ultrasoft pseudopotentials to the projector augmented-wave method},
  author = {Kresse, G. and Joubert, D.},
  journal = {Phys. Rev. B},
  volume = {59},
  issue = {3},
  pages = {1758--1775},
  numpages = {0},
  year = {1999},
  month = {Jan},
  publisher = {American Physical Society},
  doi = {10.1103/PhysRevB.59.1758},
  url = {https://link.aps.org/doi/10.1103/PhysRevB.59.1758}
}

@article{Liechtenstein1995_dftu,
  title = {Density-functional theory and strong interactions: Orbital ordering in Mott-Hubbard insulators},
  author = {Liechtenstein, A. I. and Anisimov, V. I. and Zaanen, J.},
  journal = {Phys. Rev. B},
  volume = {52},
  issue = {8},
  pages = {R5467--R5470},
  numpages = {0},
  year = {1995},
  month = {Aug},
  publisher = {American Physical Society},
  doi = {10.1103/PhysRevB.52.R5467},
  url = {https://link.aps.org/doi/10.1103/PhysRevB.52.R5467}
}

@article{Haga2022,
doi = {10.1088/1361-648X/ac5201},
url = {https://dx.doi.org/10.1088/1361-648X/ac5201},
year = {2022},
month = {feb},
publisher = {IOP Publishing},
volume = {34},
number = {17},
pages = {175601},
author = {Haga, Y and Opletal, P and Tokiwa, Y and Yamamoto, E and Tokunaga, Y and Kambe, S and Sakai, H},
title = {Effect of uranium deficiency on normal and superconducting properties in unconventional superconductor UTe2},
journal = {Journal of Physics: Condensed Matter},
}

@article{Sakai2022,
  title = {Single crystal growth of superconducting ${\mathrm{UTe}}_{2}$ by molten salt flux method},
  author = {Sakai, H. and Opletal, P. and Tokiwa, Y. and Yamamoto, E. and Tokunaga, Y. and Kambe, S. and Haga, Y.},
  journal = {Phys. Rev. Mater.},
  volume = {6},
  issue = {7},
  pages = {073401},
  numpages = {10},
  year = {2022},
  month = {Jul},
  publisher = {American Physical Society},
  doi = {10.1103/PhysRevMaterials.6.073401},
  url = {https://link.aps.org/doi/10.1103/PhysRevMaterials.6.073401}
}

@article{Braithwaite2019,
	author = {Braithwaite, D. and Vali{\v s}ka, M. and Knebel, G. and Lapertot, G. and Brison, J. -P. and Pourret, A. and Zhitomirsky, M. E. and Flouquet, J. and Honda, F. and Aoki, D.},
	da = {2019/11/22},
	doi = {10.1038/s42005-019-0248-z},
	id = {Braithwaite2019},
	isbn = {2399-3650},
	journal = {Communications Physics},
	number = {1},
	pages = {147},
	title = {Multiple superconducting phases in a nearly ferromagnetic system},
	ty = {JOUR},
	url = {https://doi.org/10.1038/s42005-019-0248-z},
	volume = {2},
	year = {2019}
}

@article{Valiska2021,
  title = {Magnetic reshuffling and feedback on superconductivity in ${\mathrm{UTe}}_{2}$ under pressure},
  author = {Vali\ifmmode \check{s}\else \v{s}\fi{}ka, M. and Knafo, W. and Knebel, G. and Lapertot, G. and Aoki, D. and Braithwaite, D.},
  journal = {Phys. Rev. B},
  volume = {104},
  issue = {21},
  pages = {214507},
  numpages = {11},
  year = {2021},
  month = {Dec},
  publisher = {American Physical Society},
  doi = {10.1103/PhysRevB.104.214507},
  url = {https://link.aps.org/doi/10.1103/PhysRevB.104.214507}
}

@article{Aoki2021,
author = {Aoki ,Dai and Kimata ,Motoi and Sato ,Yoshiki J. and Knebel ,Georg and Honda ,Fuminori and Nakamura ,Ai and Li ,Dexin and Homma ,Yoshiya and Shimizu ,Yusei and Knafo ,William and Braithwaite ,Daniel and Vali\v{s}ka ,Michal and Pourret ,Alexandre and Brison ,Jean-Pascal and Flouquet ,Jacques},
title = {Field-Induced Superconductivity near the Superconducting Critical Pressure in UTe2},
journal = {Journal of the Physical Society of Japan},
volume = {90},
number = {7},
pages = {074705},
year = {2021},
doi = {10.7566/JPSJ.90.074705},
URL = {https://doi.org/10.7566/JPSJ.90.074705},
}

@article{Honda2023,
author = {Honda ,Fuminori and Kobayashi ,Shintaro and Kawamura ,Naomi and Kawaguchi ,Saori I. and Koizumi ,Takatsugu and Sato ,Yoshiki J. and Homma ,Yoshiya and Ishimatsu ,Naoki and Gouchi ,Jun and Uwatoko ,Yoshiya and Harima ,Hisatomo and Flouquet ,Jacques and Aoki ,Dai},
title = {Pressure-induced Structural Phase Transition and New Superconducting Phase in UTe2},
journal = {Journal of the Physical Society of Japan},
volume = {92},
number = {4},
pages = {044702},
year = {2023},
doi = {10.7566/JPSJ.92.044702},
URL = {https://doi.org/10.7566/JPSJ.92.044702},
}

@ARTICLE{Lewin2023,
  title    = "A review of {UTe2at} high magnetic fields",
  author   = "Lewin, Sylvia K and Frank, Corey E and Ran, Sheng and Paglione,
              Johnpierre and Butch, Nicholas P",
  journal  = "Rep. Prog. Phys.",
  publisher = "IOP Publishing",
  volume   =  86,
  number   =  11,
  pages    = 114501,
  month    =  oct,
  year     =  2023,
  doi = {10.1088/1361-6633/acfb93},
  url = {https://dx.doi.org/10.1088/1361-6633/acfb93},
  language = "en"
}

@article{Ishizuka2021,
  title = {Periodic Anderson model for magnetism and superconductivity in ${\mathrm{UTe}}_{2}$},
  author = {Ishizuka, Jun and Yanase, Youichi},
  journal = {Phys. Rev. B},
  volume = {103},
  issue = {9},
  pages = {094504},
  numpages = {9},
  year = {2021},
  month = {Mar},
  publisher = {American Physical Society},
  doi = {10.1103/PhysRevB.103.094504},
  url = {https://link.aps.org/doi/10.1103/PhysRevB.103.094504}
}

@article{Ishizuka2019,
  title = {Insulator-Metal Transition and Topological Superconductivity in ${\mathrm{UTe}}_{2}$ from a First-Principles Calculation},
  author = {Ishizuka, Jun and Sumita, Shuntaro and Daido, Akito and Yanase, Youichi},
  journal = {Phys. Rev. Lett.},
  volume = {123},
  issue = {21},
  pages = {217001},
  numpages = {6},
  year = {2019},
  month = {Nov},
  publisher = {American Physical Society},
  doi = {10.1103/PhysRevLett.123.217001},
  url = {https://link.aps.org/doi/10.1103/PhysRevLett.123.217001}
}

@article{Harima2020,
  title     = "How to obtain Fermi surfaces of {UTe2}",
  author    = "Harima, Hisatomo",
  journal   = "JPS Conf. Proc.",
  publisher = "Journal of the Physical Society of Japan",
  volume    =  29,
  pages     = 011006,
  month     =  feb,
  year      =  2020,
  doi = {10.7566/JPSCP.29.011006},
}

@article{Hakuno2024,
  title = {Magnetism and superconductivity in mixed-dimensional periodic Anderson model for ${\mathrm{UTe}}_{2}$},
  author = {Hakuno, Ryuji and Nogaki, Kosuke and Yanase, Youichi},
  journal = {Phys. Rev. B},
  volume = {109},
  issue = {10},
  pages = {104509},
  numpages = {7},
  year = {2024},
  month = {Mar},
  publisher = {American Physical Society},
  doi = {10.1103/PhysRevB.109.104509},
  url = {https://link.aps.org/doi/10.1103/PhysRevB.109.104509}
}

@article{Ran2019,
    author = {Sheng Ran  and Chris Eckberg  and Qing-Ping Ding  and Yuji Furukawa  and Tristin Metz  and Shanta R. Saha  and I-Lin Liu  and Mark Zic  and Hyunsoo Kim  and Johnpierre Paglione  and Nicholas P. Butch },
    title = {Nearly ferromagnetic spin-triplet superconductivity},
    journal = {Science},
    volume = {365},
    number = {6454},
    pages = {684-687},
    year = {2019},
    doi = {10.1126/science.aav8645},
    URL = {https://www.science.org/doi/abs/10.1126/science.aav8645},
    eprint = {https://www.science.org/doi/pdf/10.1126/science.aav8645},
}

@article{Tokunaga2019,
  author = {Tokunaga ,Yo and Sakai ,Hironori and Kambe ,Shinsaku and Hattori ,Taisuke and Higa ,Nonoka and Nakamine ,Genki and Kitagawa ,Shunsaku and Ishida ,Kenji and Nakamura ,Ai and Shimizu ,Yusei and Homma ,Yoshiya and Li ,DeXin and Honda ,Fuminori and Aoki ,Dai},
  title = {125Te-NMR Study on a Single Crystal of Heavy Fermion Superconductor UTe2},
  journal = {J. Phys. Soc. Jpn.},
  volume = {88},
  number = {7},
  pages = {073701},
  year = {2019},
  doi = {10.7566/JPSJ.88.073701},
  URL = {https://doi.org/10.7566/JPSJ.88.073701},
  eprint = {https://doi.org/10.7566/JPSJ.88.073701},
}

@article{Sundar2019,
  title = {Coexistence of ferromagnetic fluctuations and superconductivity in the actinide superconductor ${\mathrm{UTe}}_{2}$},
  author = {Sundar, Shyam and Gheidi, S. and Akintola, K. and C\^ot\'e, A. M. and Dunsiger, S. R. and Ran, S. and Butch, N. P. and Saha, S. R. and Paglione, J. and Sonier, J. E.},
  journal = {Phys. Rev. B},
  volume = {100},
  issue = {14},
  pages = {140502},
  numpages = {5},
  year = {2019},
  month = {Oct},
  publisher = {American Physical Society},
  doi = {10.1103/PhysRevB.100.140502},
  url = {https://link.aps.org/doi/10.1103/PhysRevB.100.140502}
}

@article{Duan2020,
  title = {Incommensurate Spin Fluctuations in the Spin-Triplet Superconductor Candidate ${\mathrm{UTe}}_{2}$},
  author = {Duan, Chunruo and Sasmal, Kalyan and Maple, M. Brian and Podlesnyak, Andrey and Zhu, Jian-Xin and Si, Qimiao and Dai, Pengcheng},
  journal = {Phys. Rev. Lett.},
  volume = {125},
  issue = {23},
  pages = {237003},
  numpages = {6},
  year = {2020},
  month = {Dec},
  publisher = {American Physical Society},
  doi = {10.1103/PhysRevLett.125.237003},
  url = {https://link.aps.org/doi/10.1103/PhysRevLett.125.237003}
}

@article{Duan2021,
    author = {Duan, Chunruo and Baumbach, R. E. and Podlesnyak, Andrey and Deng, Yuhang and Moir, Camilla and Breindel, Alexander J. and Maple, M. Brian and Nica, E. M. and Si, Qimiao and Dai, Pengcheng},
    da = {2021/12/01},
    doi = {10.1038/s41586-021-04151-5},
    id = {Duan2021},
    isbn = {1476-4687},
    journal = {Nature},
    number = {7890},
    pages = {636--640},
    title = {Resonance from antiferromagnetic spin fluctuations for superconductivity in UTe2},
    ty = {JOUR},
    url = {https://doi.org/10.1038/s41586-021-04151-5},
    volume = {600},
    year = {2021}
}

@article{Raymond2021,
author = {Raymond ,St\'{e}phane and Knafo ,William and Knebel ,Georg and Kaneko ,Koji and Brison ,Jean-Pascal and Flouquet ,Jacques and Aoki ,Dai and Lapertot ,G\'{e}rard},
title = {Feedback of Superconductivity on the Magnetic Excitation Spectrum of UTe2},
journal = {J. Phys. Soc. Jpn.},
volume = {90},
number = {11},
pages = {113706},
year = {2021},
doi = {10.7566/JPSJ.90.113706},
URL = {https://doi.org/10.7566/JPSJ.90.113706},
eprint = {https://doi.org/10.7566/JPSJ.90.113706},
}

@article{Knafo2021,
  title = {Low-dimensional antiferromagnetic fluctuations in the heavy-fermion paramagnetic ladder compound ${\mathrm{UTe}}_{2}$},
  author = {Knafo, W. and Knebel, G. and Steffens, P. and Kaneko, K. and Rosuel, A. and Brison, J.-P. and Flouquet, J. and Aoki, D. and Lapertot, G. and Raymond, S.},
  journal = {Phys. Rev. B},
  volume = {104},
  issue = {10},
  pages = {L100409},
  numpages = {6},
  year = {2021},
  month = {Sep},
  publisher = {American Physical Society},
  doi = {10.1103/PhysRevB.104.L100409},
  url = {https://link.aps.org/doi/10.1103/PhysRevB.104.L100409}
}

@article{Girod2022,
  title = {Thermodynamic and electrical transport properties of ${\mathrm{UTe}}_{2}$ under uniaxial stress},
  author = {Girod, Cl\'ement and Stevens, Callum R. and Huxley, Andrew and Bauer, Eric D. and Santos, Frederico B. and Thompson, Joe D. and Fernandes, Rafael M. and Zhu, Jian-Xin and Ronning, Filip and Rosa, Priscila F. S. and Thomas, Sean M.},
  journal = {Phys. Rev. B},
  volume = {106},
  issue = {12},
  pages = {L121101},
  numpages = {5},
  year = {2022},
  month = {Sep},
  publisher = {American Physical Society},
  doi = {10.1103/PhysRevB.106.L121101},
  url = {https://link.aps.org/doi/10.1103/PhysRevB.106.L121101}
}

@article{Theuss2024,
	author = {Theuss, Florian and Shragai, Avi and Grissonnanche, Ga{\"e}l and Hayes, Ian M. and Saha, Shanta R. and Eo, Yun Suk and Suarez, Alonso and Shishidou, Tatsuya and Butch, Nicholas P. and Paglione, Johnpierre and Ramshaw, B. J.},
	da = {2024/07/01},
	doi = {10.1038/s41567-024-02493-1},
	id = {Theuss2024},
	isbn = {1745-2481},
	journal = {Nature Physics},
	number = {7},
	pages = {1124--1130},
	title = {Single-component superconductivity in UTe2 at ambient pressure},
	ty = {JOUR},
	url = {https://doi.org/10.1038/s41567-024-02493-1},
	volume = {20},
	year = {2024}
}

@article{Theuss2024_elasticmoduli,
  title = {Resonant Ultrasound Spectroscopy for Irregularly Shaped Samples and Its Application to Uranium Ditelluride},
  author = {Theuss, Florian and Simarro, Gregorio de la Fuente and Shragai, Avi and Grissonnanche, Gael and Hayes, Ian M. and Saha, Shanta and Shishidou, Tatsuya and Chen, Taishi and Nakatsuji, Satoru and Ran, Sheng and Weinert, Michael and Butch, Nicholas P. and Paglione, Johnpierre and Ramshaw, B. J.},
  journal = {Phys. Rev. Lett.},
  volume = {132},
  issue = {6},
  pages = {066003},
  numpages = {7},
  year = {2024},
  month = {Feb},
  publisher = {American Physical Society},
  doi = {10.1103/PhysRevLett.132.066003},
  url = {https://link.aps.org/doi/10.1103/PhysRevLett.132.066003}
}

@article{Vasina2024,
  title = {Connecting High-Field and High-Pressure Superconductivity in ${\mathrm{UTe}}_{2}$},
  author = {Vasina, T. and Aoki, D. and Miyake, A. and Seyfarth, G. and Pourret, A. and Marcenat, C. and Amano Patino, M. and Lapertot, G. and Flouquet, J. and Brison, J.-P. and Braithwaite, D. and Knebel, G.},
  journal = {Phys. Rev. Lett.},
  volume = {134},
  issue = {9},
  pages = {096501},
  numpages = {6},
  year = {2025},
  month = {Mar},
  publisher = {American Physical Society},
  doi = {10.1103/PhysRevLett.134.096501},
  url = {https://link.aps.org/doi/10.1103/PhysRevLett.134.096501}
}

\end{document}